\documentclass[a4paper,eqsecnum,notitlepage,nofootinbib,longbibliography]{revtex4-1}
\usepackage{slashed,bbold,bm,amsmath,amssymb,mathtools,wrapfig} 
\usepackage{color,epsfig}
\usepackage{hyperref}
\hypersetup{
    bookmarks=true,         
    unicode=false,          
    pdftoolbar=true,        
    pdfmenubar=true,        
    pdffitwindow=false,     
    pdfstartview={FitH},    
    pdftitle={My title},    
    pdfauthor={Author},     
    pdfsubject={Subject},   
    pdfcreator={Creator},   
    pdfproducer={Producer}, 
    pdfkeywords={keyword1, key2, key3}, 
    pdfnewwindow=true,      
    colorlinks=true,       
    linkcolor=red,          
    citecolor=magenta,        
    filecolor=green,      
    urlcolor=blue           
}

\newcommand{\crt}{\\[2mm]}
\newcommand{\nn}{\nonumber}
\newcommand{\beq} {\begin{equation}}
\newcommand{\eeq} {\end{equation}}
\newcommand{\beqa} {\begin{eqnarray}}
\newcommand{\eeqa} {\end{eqnarray}}

\newcommand{\cf}{{\it cf.}}
\newcommand{\ie}{{\it i.e.}}

\newcommand{\eg}{{\it e.g.}}

\newcommand{\as}{{\alpha_s}}

\newcommand{\la}{\Lambda}

\newcommand{\veps}{\varepsilon}
\newcommand{\order}[1]{${\mathcal O}\left(#1 \right)$}
\newcommand{\morder}[1]{{\mathcal O}\left(#1 \right)}
\newcommand{\eq}[1]{(\ref{#1})}
\newcommand{\fig}[1]{Fig.~\ref{#1}}
\newcommand{\lsim}{\alt}

\newcommand{\inv}[1]{\frac{1}{#1}}
\newcommand{\halft}{{\textstyle \frac{1}{2}}}

\newcommand{\quart}{{\textstyle \frac{1}{4}}}

\newcommand{\intt}{{\textstyle \int}}
\newcommand{\ket}[1]{\left\vert{#1}\right\rangle}
\newcommand{\bra}[1]{\langle{#1}\vert}

\newcommand{\com}[2]{\left[{#1},{#2}\right]}
\newcommand{\comb}[2]{\big[{#1},{#2}\big]}
\newcommand{\acom}[2]{\left\{{#1},{#2}\right\}}
\newcommand{\acomb}[2]{\big\{{#1},{#2}\big\}}
\newcommand{\tr}{\mathrm{Tr}\,}

\newcommand{\bs}[1]{\boldsymbol{#1}}

\newcommand{\bC}{\mathbb{C}}

\newcommand{\mE}{\mathcal{E}}
\newcommand{\mH}{\mathcal{H}}

\newcommand{\mJ}{\mathcal{J}}
\newcommand{\mK}{\mathcal{K}}
\newcommand{\mL}{\mathcal{L}}
\newcommand{\mM}{\mathcal{M}}
\newcommand{\mP}{\mathcal{P}}
\newcommand{\bP}{\mathbb{P}}
\newcommand{\mS}{\mathcal{S}}

\newcommand{\xv}{{\bs{x}}} 
\newcommand{\yv}{{\bs{y}}}
\newcommand{\zv}{{\bs{z}}}
\newcommand{\pv}{{\bs{p}}}
\newcommand{\kv}{{\bs{k}}}

\newcommand{\Av}{{\bs{A}}}

\newcommand{\Ev}{{\bs{E}}}
\newcommand{\Jv}{{\bs{J}}}
\newcommand{\Lv}{{\bs{L}}}
\newcommand{\Pv}{{\bs{P}}}
\newcommand{\Sv}{{\bs{S}}}
\newcommand{\ellv}{{\bs{\ell}}}
\newcommand{\gv}{\bs{\gamma}}

\newcommand{\gz}{\gamma^0}

\newcommand{\gf}{\gamma_5}

\newcommand{\nv}{\bs{\nabla}}

\newcommand{\rar}{\rightarrow}
\newcommand{\lar}{\leftarrow}

\newcommand{\rder}{{\buildrel\rar\over{\partial}}}
\newcommand{\lder}{{\buildrel\lar\over{\partial}}}

\newcommand{\rnab}{{\overset{\rar}{\nv}}\strut}
\newcommand{\lnab}{{\overset{\lar}{\nv}}\strut}
\newcommand{\rorb}{{\overset{\rar}{\bs{L}}}\strut}
\newcommand{\lorb}{{\overset{\lar}{\bs{L}}}\strut}
\newcommand{\rla}{{\overset{\rar}{\la}}\strut}  
\newcommand{\lla}{{\overset{\lar}{\la}}\strut}  

\newcommand{\xtr}{\bs{x}^\perp}

\newcommand{\atr}{{\bs{\alpha}^\perp}}
\newcommand{\alv}{{\bs{\alpha}}}

\newcommand{\aly}{{\alpha^2}}
\newcommand{\alz}{{\alpha^3}}

\newcommand{\wfp}{\Phi^{(\Pv)}}
\newcommand{\wfr}{{\Phi^{(0)}}}

\newcommand{\sph}{{Y_{j\lambda}}}

\begin{document}

\title{Bound states and QCD}\thanks{Published exclusively as arXiv:1807.05598v2.}

\author{Paul Hoyer}
\affiliation{Department of Physics, POB 64, FIN-00014 University of Helsinki, Finland}

\begin{abstract} 

The similarities of hadrons and atoms motivate a study of the principles of QED bound states and of their applicability to QCD. The power series in $\alpha$ and $\log\alpha$ of the binding energy is reflected in the Fock expansion of the bound state in temporal gauge ($A^0=0$). Gauss' constraint on physical states fixes the gauge for time independent transformations and determines the instantaneous interaction within each Fock state.

\vspace{.1cm}

Positronium atoms generate a classical (dipole) electric field, whereas there can be no color octet gluon field for color singlet hadrons. Hence the gluon field generated by each color component of a hadron need not vanish at spatial infinity. Gauss' constraint has a homogeneous solution with a single parameter $\Lambda$ that is compatible with Poincar\'e invariance. The corresponding potential is linear for $q\bar q$ and $gg$ Fock states, and confining also for other states ($q\bar qg,\,qqq$).

\vspace{.1cm}

This approach is consistent with the quarkonium phenomenology based on the Cornell potential at lowest order. The relativistic meson and glueball eigenstates of the QCD Hamiltonian with the \order{\alpha_s^0} linear potential are determined. The states lie on linear Regge trajectories and their daughters. There are also massless bound states which allow to include a $J^{PC}=0^{++}$ condensate in the perturbative vacuum, thus breaking chiral symmetry spontaneously.

\end{abstract}

\maketitle
\vspace{-.5cm}

\tableofcontents

\parindent 0cm
\vspace{-.2cm}

\section{Introduction \label{sI}}

Hadrons and atoms have common features. This is immediately apparent for heavy quarkonia, whose spectra and decays are well described by the Schr\"odinger equation with the Cornell potential \cite{Eichten:1979ms,Eichten:2007qx},
\begin{align} \label{eI1}
V(r) = V'r-\frac{4}{3}\frac{\as}{r} \ \ \ \text{with}\ \ V' \simeq 0.18\ \text{GeV}^2, \ \ \as \simeq 0.39
\end{align}
This potential, based on fits to data, was later found to agree with Lattice QCD \cite{Bali:2000gf}. Color confinement is realized with a linear potential and the gluon coupling $\as$ is close to $\as(m_\tau) \simeq 0.33$. 

This and other features of hadrons (see, \eg, \cite{Dokshitzer:1998nz,Dokshitzer:2003bt,Dokshitzer:2010zza,Hoyer:2016aew}) suggest perturbative aspects of QCD bound states. It may seem unlikely that a perturbative approach akin to that for QED atoms could describe confinement and chiral symmetry breaking. However, bound state perturbation theory differs from that for scattering amplitudes. 

Feynman diagrams do not have bound state poles, even in QED. The residues are given by the atomic wave functions, which are non-polynomial in $\alpha$. The perturbative expansion of a bound state wave function is not unique, as first recognized for the Bethe-Salpeter equation \cite{Caswell:1978mt,Lepage:1978hz}. Wave functions are not directly observable, and typically gauge dependent. Binding energies on the other hand are measurable, and do have a unique perturbative expansion in $\alpha$ and $\log\alpha$. Positronium calculations are compared with data in \cite{Penin:2014bea,Adkins:2015wya}.

The expansion of scattering amplitudes in terms of Feynman diagrams is derived in the Interaction Picture (IP). It is formally exact if the true initial and final states have a non-vanishing overlap with the free $in$ and $out$ states of the IP. Bound states have finite size and thus zero overlap with non-interacting states. The free propagators of Feynman diagrams are inappropriate for bound states, whose constituents move in a non-vanishing field (see chapter 14 of \cite{Weinberg:1995mt}).

The $\ket{e^+e^-}$ Fock state of Positronium suffices (in the rest frame) to determine the binding energy at leading order, $E_b = -\quart m_e\alpha^2$. The wave function of the $\ket{e^+e^-}$ component satisfies the Schr\"odinger equation and is exponential in $\alpha$. Fock states with more constituents, such as $\ket{e^+e^-\gamma}$, contribute to $E_b$ at \order{\alpha^4} and higher. In this way the well-defined perturbative expansion of the binding energy is mirrored in the Fock state expansion.

The successful phenomenology based on \eq{eI1} indicates that quarkonia are dominated by Fock states with two heavy quarks. This is non-trivial, as their binding energies are large compared to the masses of light quarks and gluons. The $q\bar q$ and $qqq$ quantum numbers of light hadrons similarly suggest that their simplest Fock states dominate, and that transversely polarized gluons may be treated as a perturbation.

Field theory interactions are generated by the exchange of particles. For example, the \order{\alpha^4} spin dependence of Positronium states arises from the exchange of transverse photons. The bound state acquires an $\ket{e^+e^-\gamma}$ Fock state during the propagation time of the photon, which is of \order{\alpha} compared to the lifetime of the $\ket{e^+e^-}$ component \cite{Jarvinen:2004pi}.

Gauge theories have an instantaneous $A^0$ field which generates interactions within Fock states. The $\ket{e^+e^-}$ Fock state of Positronium can dominate because it is bound by $A^0$. For an electron at $\xv_1$ and positron at $\xv_2$ Gauss' law specifies
\begin{align} \label{eI2}
-\nv^2 A^0(\xv) = e\delta(\xv-\xv_1)-e\delta(\xv-\xv_2)\ \ \ \ \ {\color{red} \Longrightarrow}\ \ \ \ \ A^0(\xv) = \frac{e}{4\pi} \Big(\frac{1}{|\xv-\xv_1|} -\frac{1}{|\xv-\xv_2|}\Big)
\end{align}

This allows to derive the Schr\"odinger equation from the QED action (section \ref{sIIA}). Positronium is approximated as an $\ket{e^+e^-}$ state where the electron and positron are distributed according to a wave function $\Phi(\xv_1,\xv_2)$. For this state to be an eigenstate of the QED Hamiltonian, with the gauge field \eq{eI2}, $\Phi(\xv_1,\xv_2)$ has to satisfy a bound state equation (BSE) with the classical potential
\begin{align} \label{eI3}
V(|\xv_1-\xv_2|) = \halft e\big[A^0(\xv_1)-A^0(\xv_2)\big] = -\frac{\alpha}{|\xv_1-\xv_2|}
\end{align}
The factor $\halft$ is due to the field energy contribution $\int d\xv\,\quart F_{\mu\nu}(\xv)F^{\mu\nu}(\xv)$, with $A^0$ as in \eq{eI2} and $\Av=0$. The ``self-energies'' $\propto 1/|\xv_1-\xv_1|,\ 1/|\xv_2-\xv_2|$ may be subtracted, since they are independent of $\xv_1,\xv_2$. The BSE reduces to the Schr\"odinger equation in the non-relativistic limit.

This straightforward derivation of the Schr\"odinger equation applies also to quarkonia in QCD. How then can the linear term in the potential \eq{eI1} arise? The confining potential should be due to $A^0$ for the $\ket{q\bar q}$ Fock state to dominate. Our only implicit assumption was the boundary condition $\lim_{|\xv|\to\infty}A^0(\xv)=0$ in solving Gauss' law \eq{eI2}. This ensured that the electric field of the atom decreases with distance.

Quarkonia are singlets of color SU(3). Hence they do not generate a color octet gluon field $A^0_a(\xv)$ analogous to \eq{eI2} at any $\xv$. The classical gluon field of each color component $\ket{q^C(\xv_1)\bar q^C(\xv_2)}$ cancels in the sum over quark colors $C$. Hence an observer who is external to the bound state finds $A^0_a(\xv)=0$ at all $\xv$. On the other hand, the quark $q^C(\xv_1)$ feels the non-vanishing gluon field of its companion $\bar q^C(\xv_2)$.

This motivates us to consider homogeneous solutions of Gauss' law, in which the gluon field of $\ket{q^C(\xv_1)\bar q^C(\xv_2)}$ for a given color $C$ does not vanish at spatial infinity (section \ref{sIIB}). Translation and rotation invariance requires the sourceless solution to have a spatially constant field energy density, characterized by a universal constant $\la$. This solution gives rise to an \order{\alpha_s^0} linear potential, with $V'=\la^2$ in \eq{eI1}.

The dominance of the classical gluon field over loop corrections implies that $\as(Q)$ stops running for $Q$ of \order{\la}, explaining the moderate size of $\as$ in \eq{eI1}. Hence we have a perturbative derivation of the successful quarkonium phenomenology. The frozen coupling $\as$ remains perturbative for light quarks and gluons. This approach can thus be applied also to light hadrons and glueballs, which are relativistically bound by the linear potential.

In the next section we demonstrate our method by deriving the Schr\"odinger equation for Positronium and the confining potential of QCD. The following sections discuss various properties of the solutions, adding some material to the first version \cite{Hoyer:2018hdj} of this paper. Section \ref{sVIII} provides a summary.

\section{Potential energy for QED and QCD states \label{sII}}

In a perturbative expansion field fluctuations are suppressed by powers of the coupling. Thus  tree diagrams dominate loop contributions to scattering amplitudes. For bound states the classical gauge field gives the leading contribution to binding (\cf\ the $-\alpha/r$ potential of Positronium). In the sense that loops bring factors of $\hbar$ perturbation theory may be thought of as an expansion in $\hbar$.

Bound states are eigenstates of the Hamiltonian and may be expanded in Fock states defined at an instant of time. The instantaneous $A^0$ field contributes to binding without being a Fock state constituent. In Positronium the dominant $\ket{e^+e^-}$ Fock state is bound by the classical $A^0$ field. Fock components with more constituents contribute to the binding energy at higher orders of its perturbative expansion.

Coulomb gauge ($\nv\cdot\Av =0$) \cite{Feinberg:1977rc} is common in bound state calculations, but we shall use the temporal gauge, $A^0 = 0$ \cite{Willemsen:1977fr,Bjorken:1979hv,Christ:1980ku,Leibbrandt:1987qv,Strocchi:2013awa}. The role of $A^0$ is then taken over by the longitudinal electric field $\Ev_L$. Gauss' law is implemented as a \textsl{constraint} on Fock states, rather than as an operator identity. $\Ev_L$ gives each Fock state an instantaneous potential energy without creating new constituents. We demonstrate this first in QED and then in QCD.  

\subsection{Positronium in temporal gauge \label{sIIA}}

The QED action
\begin{align} \label{eII1}
\mS &= \int d^4x\big[-\quart F_{\mu\nu}F^{\mu\nu}+\bar\psi(i\slashed{\partial}-m-e\slashed{A})\psi\big]
& F_{\mu\nu} = \partial_\mu A_\nu-\partial_\nu A_\mu
\end{align}
has no $\partial_0 A^0$ term, so $A^0$ lacks a conjugate field. This makes temporal gauge ($A^0=0$) convenient for canonical quantization. The electric field $E^i=F^{i0}=-\partial_0 A^i$ is conjugate to $A_i \ (i=1,2,3)$, and $i\psi_\alpha^\dag$ is conjugate to $\psi_\alpha$, giving the canonical commutation relations
\begin{align} \label{eII2}
\com{E^i(t,\xv)}{A^j(t,\yv)} &= i\delta^{ij}\delta(\xv-\yv)  & \acom{\psi^\dag_\alpha(t,\xv)}{\psi_\beta(t,\yv)} = \delta_{\alpha\beta}\,\delta(\xv-\yv)
\end{align}
The Hamiltonian
\begin{align} \label{eII3}
\mH = \int d\xv\big[E^i\partial_0 A_i+i\psi^\dag\partial_0\psi-\mL\big]
= \int d\xv\big[\halft E^iE^i +\quart F^{ij}F^{ij}+\psi^\dag(-i\alpha^i\partial_i-e\alpha^iA^i+m\gz)\psi\big] 
\end{align}
involves only the transverse gauge field $A_T^i$ (any $A_L^i$ contribution to the fermion part may be removed by a redefinition of the phase of the fermion field). There remains a contribution of the longitudinal electric field: $\Ev^2 = \Ev_L^2+\Ev_T^2$.

The Gauss' operator defined by
\begin{align} \label{eII4}
G(x) \equiv \frac{\delta\mS}{\delta{A^0(x)}} = \partial_i E^{i}(x)-e\psi^\dag\psi(x)
\end{align}
does not vanish as an operator identity in $A^0=0$ gauge. $G(x)$ is time independent since it commutes with the Hamiltonian, $\com{G(t,\xv)}{\mH(t)}=0$. Gauss' operator generates time-independent gauge transformations as follows. For an infinitesimal gauge parameter $\delta\la(\xv)$ the unitary operator
\begin{align} \label{eII5}
U(t) = 1+i\int d\yv\,G(t,\yv)\delta\la(\yv) = 1+i\int d\yv\,\big[\partial_i E^i(t,\yv)-e\psi^\dag\psi(t,\yv)\big]\delta\la(\yv)
\end{align}
implements, via the canonical commutation relations \eq{eII2}, the gauge transformations
\begin{align} \label{eII6}
&U(t)A^j(t,\xv)U^{-1}(t)= A^j(t,\xv) + \partial_j \delta\la(\xv)
&U(t)\psi(t,\xv)U^{-1}(t)= \psi(t,\xv) + ie\,\delta\la(\xv)\psi(t,\xv)
\end{align}

The temporal gauge condition $A^0=0$ is invariant under time-independent gauge transformations. The gauge is fully fixed for physical states by imposing the vanishing of the generator $G(x)$ as a constraint,
\begin{align} \label{eII7}
G(x)\ket{phys} = 0
\end{align}
This defines the action of the longitudinal electric field,
\begin{align} \label{eII8}
E_L^i(t,\xv)\ket{phys} &= -\partial_i^x \int d\yv \frac{e}{4\pi|\xv-\yv|}\psi^\dag\psi(t,\yv)\ket{phys}
\end{align}
Assuming $E_L^i\ket{0}=0$ we require for an $e^+e^-$ component of Positronium (at any time $t$),
\begin{align} \label{eII9}
E_L^i(\xv)\ket{e^-(\xv_1)e^+(\xv_2)} = \com{E_L^i(\xv)}{\bar\psi_\alpha(\xv_1)\psi_\beta(\xv_2)}\ket{0}=0
\end{align}
Corresponding to \eq{eI2} we have then,
\begin{align} \label{eII10}
E_L^i\ket{e^-(\xv_1)e^+(\xv_2)}&= -\partial_i^x \int d\yv \frac{e}{4\pi|\xv-\yv|}\com{\psi^\dag\psi(\yv)}{\bar\psi_\alpha(\xv_1)\psi_\beta(\xv_2)}\ket{0} \nn\crt
&=-\partial_i^x\, \frac{e}{4\pi} \Big(\frac{1}{|\xv-\xv_1|} -\frac{1}{|\xv-\xv_2|}\Big)\ket{e^-(\xv_1)e^+(\xv_2)}
\end{align}

With a partial integration the contribution of $E_L^i$ to the Hamiltonian \eq{eII3} becomes,
\begin{align} \label{eII11}
\mH_V\ket{phys} \equiv \halft\int d\xv E_L^i E_L^i(\xv)\ket{phys} 
&= \halft\int d\xv d\yv d\zv\Big[\partial_i^x \frac{e}{4\pi|\xv-\yv|}\psi^\dag\psi(\yv)\Big]
\Big[\partial_i^x \frac{e}{4\pi|\xv-\zv|}\psi^\dag\psi(\zv)\Big]\ket{phys} \nn\crt
&= \halft\int d\xv d\yv\, \frac{e^2}{4\pi|\xv-\yv|}\big[\psi^\dag\psi(\xv)\big]
\big[\psi^\dag\psi(\yv)\big]\ket{phys}
\end{align}
Applied to the $e^+e^-$ Fock state we get, subtracting the (infinite) self-energy contributions with $\xv=\yv$, the classical potential of \eq{eI3},
\begin{align} \label{eII12}
\mH_V\ket{e^-(\xv_1)e^+(\xv_2)}= -\frac{\alpha}{|\xv_1-\xv_2|}\ket{e^-(\xv_1)e^+(\xv_2)} \equiv V(|\xv_1-\xv_2|)\ket{e^-(\xv_1)e^+(\xv_2)}
\end{align}

We express Positronium as a superposition of $e^+e^-$ Fock states distributed according to a wave function $\Phi$, which in the rest frame is a function only of $\xv_1-\xv_2$. Denoting the Positronium mass by $M$,
\begin{align} \label{eII13}
\ket{M} = \int d\xv_1 d\xv_2\,\bar\psi_\alpha(t=0,\xv_1)\Phi_{\alpha\beta}(\xv_1-\xv_2)\psi_\beta(t=0,\xv_2)\ket{0}
\end{align}
In the bound state condition
\begin{align} \label{eII14}
\mH\ket{M} = M\ket{M}
\end{align}
we may, at leading order in $\alpha$, ignore the transverse photon created by the $e\bar\psi \,\alv\cdot\Av_T\psi$ term in the Hamiltonian \eq{eII3}. The commutators of the free fermion Hamiltonian $\mH_0^{(f)} = \intt d\xv\,\psi^\dag(\xv)(-i\alv\cdot\nv+m\gz)\psi(\xv)$ contribute
\begin{align} \label{eII14a}
\com{\mH_0^{(f)}}{\bar\psi(\xv_1)} &= \bar\psi(\xv_1)(-i\alv\cdot\lnab_1+m\gz) 
&\com{\mH_0^{(f)}}{\psi(\xv_2)} =
(i\alv\cdot\rnab_2-m\gz)\psi(\xv_2)
\end{align}
giving the bound state equation for $\Phi(\xv)$,
\begin{align} \label{eII15}
\big(i\alv\cdot\rnab_1+m\gz\big)\Phi(\xv_1-\xv_2)-\Phi(\xv_1-\xv_2)\big(i\alv\cdot\lnab_2+m\gz\big) = \big(M- V\big)\Phi(\xv_1-\xv_2)
\end{align}
where $\nv_i = \partial/\partial\xv_i$ and $V= V(|\xv_1-\xv_2|)$ is the potential defined in \eq{eII12}.

Since \eq{eII15} is valid only at leading order in $\alpha$ it may, without loss of information, be reduced to the non-relativistic Schr\"odinger equation. The Schr\"odinger wave function $\phi(\xv)$ is a scalar function that is independent of the electron and positron helicities. It defines the Positronium state \eq{eII13} as
\begin{align} \label{eII16}
\ket{M} = \int\frac{d\kv}{(2\pi)^3}\,\phi(\kv)\,b^\dag_{\kv,\lambda_1}d^\dag_{-\kv,\lambda_2}\ket{0} = \int\frac{d\kv}{(2\pi)^3} d\xv\,\phi(\xv)e^{-i\kv\cdot\xv}\,b^\dag_{\kv,\lambda_1}d^\dag_{-\kv,\lambda_2}\ket{0}
\end{align}
where $\phi(\kv)$ denotes the Fourier transform of $\phi(\xv)$. The fermion fields in \eq{eII13} are defined as
\begin{align} \label{eII17}
\psi_\alpha(t=0,\xv) = \int\frac{d\kv}{(2\pi)^3 2E_k}\sum_\lambda\big[u_\alpha(\kv,\lambda)e^{i\,\kv\cdot\xv}b_{\kv,\lambda}+v_\alpha(\kv,\lambda)e^{-i\,\kv\cdot\xv}d_{\kv,\lambda}^\dag\big]
\end{align}
Only the $b^\dag$ operator in $\bar\psi(\xv_1)$ and $d^\dag$ in $\psi(\xv_2)$ contribute in  in the non-relativistic limit. Comparing \eq{eII16} with \eq{eII13} allows to express the $4\times 4$ wave function $\Phi_{\alpha\beta}$ in terms of the Schr\"odinger wave function, 
\begin{align} \label{eII18}
\Phi_{\alpha\beta}(\xv) = {_\alpha\big[}\gz u(-i\,\rnab,\lambda_1)\big]\,\phi(\xv)\big[\bar v(i\,\lnab,\lambda_2)\gz\big]_\beta
\end{align}

The bound state equation \eq{eII15} may be expressed as ($\xv=\xv_1-\xv_2$),
\begin{align}\label{eII19}
\Big[\frac{2}{M-V}(i\alv\cdot\rnab+m\gz)-1\Big]\Phi(\xv)+\Phi(\xv)\Big[(i\alv\cdot\lnab-m\gz)\frac{2}{M-V}-1\Big]=0
\end{align}
The potential $V$ as well as the binding energy $E_b \equiv M-2m$ are of \order{\alpha^2} so we may expand,
\begin{align} \label{eII20}
\frac{2}{M-V} \simeq \inv{E}+\inv{2m^2}\Big[-\frac{\nv^2}{m}+V-E_b\Big]
\end{align}
where we used $E\equiv\sqrt{-\nv^2+m^2} \simeq m-\nv^2/2m$. The expression \eq{eII18} of $\Phi_{\alpha\beta}$ and the properties of the $u$ and $v$ spinors imply
\begin{align} \label{eII21}
\Big[\inv{E}(i\alv\cdot\rnab+m\gz)-1\Big]\Phi(\xv) = \Phi(\xv)\Big[(i\alv\cdot\lnab-m\gz)\inv{E}-1\Big] = 0
\end{align}
so the term $1/E$ in \eq{eII20} does not contribute to the BSE \eq{eII19}. The second term in \eq{eII20} can be brought past the derivatives in the expression \eq{eII18} for $\Phi$, since $\nv V(\xv)$ is of \order{\alpha^3} and may be ignored. Both terms in the BSE \eq{eII19} then vanish separately, given that $\phi(\xv)$ satisfies the Schr\"odinger equation (with reduced mass $\halft m$),
\begin{align} \label{eII22}
\Big[-\frac{\nv^2}{m}+V\Big]\phi(\xv) = E_b \phi(\xv)
\end{align}

This completes the derivation of the Schr\"odinger equation in the temporal gauge of QED. It will obviously be important to check that this method gives the correct expression for the binding energy also at higher orders in $\alpha$.

\subsection{Potential in QCD \label{sIIB}}

We consider the QCD action
\begin{align} \label{eII23}
\mS &= \int d^4x\big[-\quart F_{\mu\nu}^aF^{\mu\nu}_a+\bar\psi(i\slashed{\partial}-m-g\slashed{A}_a T^a)\psi\big]
& F_{\mu\nu}^a = \partial_\mu A_\nu^a-\partial_\nu A_\mu^a-gf_{abc}A_\mu^bA_\nu^c
\end{align}
in temporal gauge, $A^0_a=0$ \cite{Willemsen:1977fr,Bjorken:1979hv,Christ:1980ku,Leibbrandt:1987qv}. The electric field $E_a^i=F_a^{i0}=-\partial_0 A_a^i$ is conjugate to $A_i^a=-A_a^i$, giving the equal-time commutation relations
\begin{align} \label{eII24}
\com{E_a^i(t,\xv)}{A_b^j(t,\yv)} &= i\delta_{ab}\delta^{ij}\delta(\xv-\yv)  & \acom{\psi^{A\,\dag}_\alpha(t,\xv)}{\psi_\beta^B(t,\yv)} = \delta^{AB}\delta_{\alpha\beta}\,\delta(\xv-\yv)
\end{align}
and the Hamiltonian
\begin{align} \label{eII25}
\mH &= \int d\xv\big[E_a^i\partial_0 A_i^a+i\psi^\dag\partial_0\psi-\mL\big]
= \int d\xv\big[\halft E_a^iE_a^i +\quart F_a^{ij}F_a^{ij}+\psi^\dag(-i\alv\cdot\nv+m\gz-g\alv\cdot \Av_a T^a)\psi\big]
\end{align}
where
\begin{align} \label{eII26}
\int d\xv\, \quart F^{ij}F^{ij} = \int d\xv\big[ \halft A_a^i(-\delta_{ij}\nv^2+\partial_i\partial_j)A_a^j+ gf_{abc}(\partial_i A_a^j)A_b^i A_c^j+ \quart g^2 f_{abc}f_{ade}A_b^i A_c^j A_d^i A_e^j \big]
\end{align}
contains both longitudinal and transverse gluon fields. As in QED, $\Av_L$ may be removed from the fermion part of $\mH$ by redefining the phase of the $\psi$ field.

Gauss' operator
\begin{align} \label{eII27}
G_a(x) \equiv \frac{\delta\mS}{\delta{A_a^0(x)}} = \partial_i E_a^{i}(x)+g f_{abc}A_b^i E_c^i-g\psi^\dag T^a\psi(x)
\end{align}
generates time-independent gauge transformations similarly as in QED \eq{eII5}, which leave the gauge condition $A^0_a=0$ invariant. The gauge may be fixed by constraining physical states to satisfy
\begin{align} \label{eII28}
G_a(x)\ket{phys} = 0
\end{align}
This constraint is independent of time since Gauss' operator commutes with the Hamiltonian, $\com{G_a(t,\xv)}{\mH(t)} = 0$.
Eq. \eq{eII28} has multiple solutions for non-perturbative gauge fields, the so-called ``Gribov copies'' \cite{Gribov:1977wm}. The copies do not contribute to a perturbative expansion, however. 

Gauss' constraint defines the action of the longitudinal electric field on physical states,
\begin{align} \label{eII29}
\partial_i E_{L,a}^{i}(\xv)\ket{phys} = g\big[- f_{abc}A_b^i E_c^i+\psi^\dag T^a\psi(\xv)\big]\ket{phys}
\end{align}
At higher orders in $g$ one needs to take into account the contribution of $\Ev_L$ on the rhs. We ignore this here. Then we may solve for $E_{L,a}^{i}$ as in \eq{eII8}. This is where an opportunity arises for introducing a confining potential. A quarkonium state analagous to Positronium \eq{eII13} has a color diagonal wave function,
\begin{align} \label{eII30}
\ket{M} = \inv{\sqrt{N_C}}\sum_{A,B;\alpha,\beta}\int d\xv_1 d\xv_2\,\bar\psi_\alpha^A(t=0,\xv_1)\delta^{AB}\Phi_{\alpha\beta}(\xv_1-\xv_2)\psi_\beta^B(t=0,\xv_2)\ket{0}
\end{align}
where $N_C=3$ is the number of colors. $\ket{M}$ is a color singlet under global gauge transformations\footnote{As emphasized in \cite{Strocchi:2013awa}, global and local transformations should be distinguished.}, and cannot create a classical color octet field at any position $\xv$. A single quark color component $\ket{q^C\bar q^C}$ does have an octet field, but it cancels in the sum over $C$. The unobservable gluon field of the $\ket{q^C\bar q^C}$ component need not vanish at spatial infinity. This allows also homogeneous solutions for $\Ev_L$ in \eq{eII29}, which satisfy $\partial_iE_{L,a}^{i}=0$. Specifically, we consider
\begin{align} \label{eII31}
E^i_{L,a}(\xv)\ket{phys} &= -\partial_i^x \int d\yv \Big[\kappa\,\xv\cdot\yv + \frac{g}{4\pi|\xv-\yv|}\Big]\mE_a(\yv) \ket{phys}\nn\crt
\mE_a(\yv) &= - f_{abc}A_b^i E_c^i(\yv)+\psi^\dag T^a\psi(\yv)
\end{align}
with a normalization $\kappa$ that is independent of $\xv$ and $\yv$. Translation invariance requires that the homogeneous (sourceless) contribution to $E^i_{L,a}$ is $\xv$-independent. This imposes the linear dependence on $\xv$ in the term $\propto \kappa$.

The contribution of the longitudinal electric field to the Hamiltonian \eq{eII25} is
\begin{align} \label{eII32}
\mH_V &\equiv \halft\int d\xv\,E_{a,L}^i E_{a,L}^i = \halft\int d\xv \Big\{\partial_i^x \int d\yv\Big[\kappa\, \xv\cdot\yv+\frac{g}{4\pi|\xv-\yv|}\Big]\mE_a(\yv)\Big\}
\Big\{\partial_i^x \int d\zv\Big[\kappa\, \xv\cdot\zv+\frac{g}{4\pi|\xv-\zv|}\Big]\mE_a(\zv)\Big\} \nn\crt
&= \int d\yv d\zv\Big\{\,\yv\cdot\zv \Big[\halft\kappa^2\intt d\xv + g\kappa\Big] + \halft \frac{\as}{|\yv-\zv|}\Big\}\mE_a(\yv)\mE_a(\zv) \equiv \Big[\halft\kappa^2\intt d\xv + g\kappa\Big]h_V^{(0)}+\halft\as h_V^{(1)}
\equiv \mH_V^{(0)} + \mH_V^{(1)}
\end{align} 
where the terms of \order{g\kappa,g^2} were integrated by parts. The \order{\kappa^2} term is proportional to the volume of space because of the $\xv$-independent field energy density. This term may be subtracted provided it is the same for all Fock components of all bound states. This determines the homogeneous contribution up to a universal scale $\la$. We now demonstrate this by considering several examples.

\subsubsection{$q\bar q$ states \label{sIIB1}}

Consider a component of the $q\bar q$ (meson) bound state state \eq{eII30},
\begin{align} \label{eII33}
\ket{q(\xv_1)\bar q(\xv_2)} \equiv \bar\psi_\alpha^{A}(\xv_1)\,\psi^{A}_\beta(\xv_2)\ket{0}
\end{align}
Sums over repeated color indices (here $\sum_A$) will be understood throughout, and the (fixed) Dirac indices $\alpha,\,\beta$ will be suppressed. The vacuum is assumed to satisfy $E^i_{L,a}\ket{0}=0$. With the definition of $h_V^{(0)}$ in \eq{eII32}, the action of $\mE_a(\yv)\mE_a(\zv)$ on the quark fields in \eq{eII33} contribute terms $\propto \yv\cdot\zv$, where $\yv$, $\zv$ are $\xv_1$ or $\xv_2$,
\begin{align} \label{eII34}
h_V^{(0)} \ket{q(\xv_1)\bar q(\xv_2)} &= \int d\yv d\zv\;\yv\cdot\zv\,\mE_a(\yv)\mE_a(\zv) \ket{q(\xv_1)\bar q(\xv_2)} = (\xv_1^2+\xv_2^2-2\xv_1\cdot\xv_2)\bar\psi_{A}(\xv_1)\,T^a_{AB}T^a_{BC}\psi_{C}(\xv_2)\ket{0} \nn\crt
&= C_F\,(\xv_1-\xv_2)^2\,\ket{q(\xv_1)\bar q(\xv_2)}  \hspace{2cm} C_F = \frac{N^2-1}{2N} = \frac{4}{3}\ \ (N=3)
\end{align}
For the \order{\kappa^2} term in $\mH_V$ to be universal it must be independent of $\xv_1-\xv_2$. Hence for a $q\bar q$ Fock state we choose the normalization $\kappa$ of the homogeneous solution in \eq{eII31} to be
\begin{align} \label{eII35}
\kappa_{q\bar q} = \frac{\la^2}{gC_F}\frac{1}{|\xv_1-\xv_2|}
\end{align}
This defines the universal scale $\la$, which has the dimension of energy. Subtracting the universal, \order{\kappa^2} contribution to the eigenvalue of $\mH_V$,
\begin{align} \label{eII36a}
V^{(U)} =\frac{\la^4}{2g^2C_F}\intt d\xv
\end{align}
we have
\begin{align} \label{eII36}
\mH_V\ket{q(\xv_1)\bar q(\xv_2)} &=\big[
V_{q\bar q}^{(0)}+ V_{q\bar q}^{(1)}\big]\ket{q(\xv_1)\bar q(\xv_2)} \nn\crt
V_{q\bar q}^{(0)}(\xv_1-\xv_2) &= g\kappa_{q\bar q}\,C_F\,(\xv_1-\xv_2)^2 = \la^2|\xv_1-\xv_2|  \nn\crt
V_{q\bar q}^{(1)}(\xv_1-\xv_2) &=-C_F\frac{\as}{|\xv_1-\xv_2|}
\end{align}
The linearity of the confining potential $V_{q\bar q}^{(0)}$ is a consequence of the translation and rotation invariance of the homogeneous contribution to $\Ev_L$ in \eq{eII31}.
The \order{\as} gluon exchange potential $V^{(1)}$ arising from $\mH_V^{(1)}$ in \eq{eII32} also agrees with the Cornell potential \eq{eI1}. Including the kinetic fermion term in the QCD Hamiltonian \eq{eII25} and imposing the stationarity condition $\mH\ket{M} = M\ket{M}$ on the $q\bar q$ state \eq{eII30} gives the bound state equation \eq{eII15} for the wave function, where $V = V_{q\bar q}^{(0)}+V_{q\bar q}^{(1)}$. In the non-relativistic limit this reproduces the successful quarkonium phenomenology based on the Cornell potential. We consider the properties of the relativistic $q\bar q$ states in section \ref{sIII}.

At \order{\as^0} only the linear potential $V_{q\bar q}^{(0)}$ in \eq{eII36} and the $q\bar q$ Fock components contribute, even for light (relativistic) quarks. The $\ket{q\bar qg}$ Fock state is created at \order{g} by gluon emission from the quarks. Next we consider the instantaneous potential for those states.

\subsubsection{$q\bar qg$ states \label{sIIB2}}

The Hamiltonian \eq{eII25} creates \order{g} $\ket{q\bar qg}$ Fock states from $\ket{q\bar q}$. We consider the instantaneous potential generated by $\mH_V$ \eq{eII32} for (globally) color singlet states with a transversely polarized gluon of the form
\begin{align} \label{eII37}
\ket{qg\bar q} \equiv \bar\psi_A(\xv_1)\,A_b^j(\xv_g)T^b_{AB}\psi_B(\xv_2)\ket{0}
\end{align}
where sums over colors are understood.
Also the gluon term $-g f_{abc}A_b^j E_c^j$ in $\mE_a$ \eq{eII31} now contributes. The terms $\propto \yv\cdot\zv$ in $h_V^{(0)}\ket{qg\bar q}$ are ($N=N_C=3$)
\begin{align} \label{eII38}
\xv_1^2:& \hspace{1cm} \bar\psi_{A''}(\xv_1)\,A_b^j(\xv_g)\,\psi_B(\xv_2)\ket{0}T^a_{A''A'}T^a_{A'A}T^b_{AB} = C_F\ket{qg\bar q} \nn\crt
\xv_1\cdot\xv_2:& \hspace{1cm} -2\, \bar\psi_{A'}(\xv_1)\,A_b^j(\xv_g)\,\psi_{B'}(\xv_2)\ket{0}T^a_{A'A}T^b_{AB}T^a_{BB'} = \frac{1}{N}\ket{qg\bar q} \nn\crt
\xv_1\cdot\xv_g:& \hspace{1cm} 2\, \bar\psi_{A'}(\xv_1)\,A_b^j(\xv_g)\,\psi_{B}(\xv_2)\ket{0}T^a_{A'A}(-i)f_{abc}T^c_{BB'} = -N\ket{qg\bar q} \nn\crt
\xv_g^2:& \hspace{1cm} \bar\psi_{A}(\xv_1)\,A_b^j(\xv_g)\,\psi_{B}(\xv_2)\ket{0}f_{aeb}f_{aed}T^d_{AB} = N\ket{qg\bar q}
\end{align}
The contributions $\propto \xv_2^2$ and $\xv_2\cdot\xv_g$ equal those $\propto \xv_1^2$ and $\xv_1\cdot\xv_g$, respectively. Altogether,
\begin{align} \label{eII39}
h_V^{(0)}\ket{q(\xv_1)g(\xv_g)\bar q(\xv_2)} &= \big[d_{qgq}(\xv_1,\xv_g,\xv_2)\big]^2\ket{q(\xv_1)g(\xv_g)q(\xv_2)} \nn\crt
d_{qgq}(\xv_1,\xv_g,\xv_2) &\equiv \sqrt{\quart(N-2/N)(\xv_1-\xv_2)^2+N(\xv_g-\halft\xv_1-\halft\xv_2)^2}
\end{align}
For the \order{\kappa^2} contribution to the eigenvalue of $\mH_V$ to be universal it should equal $V^{(U)}$ of \eq{eII36a}, imposing
\begin{align} \label{eII40}
\kappa_{qgq} = \frac{\la^2}{g\sqrt{C_F}}\,\frac{1}{d_{qgq}(\xv_1,\xv_g,\xv_2)}
\end{align}
The \order{g\kappa} contribution to $\mH_V$ gives the potential,
\begin{align} \label{eII41}
V_{qgq}^{(0)}(\xv_1,\xv_g,\xv_2) = g\kappa_{qgq}\,\big[d_{qgq}(\xv_1,\xv_g,\xv_2)\big]^2 = \frac{\la^2}{\sqrt{C_F}}\, d_{qgq}(\xv_1,\xv_g,\xv_2)
\end{align}
When the gluon coincides with one of the quarks the potential $V_{qgq}^{(0)}(\xv_1=\xv_g,\xv_2) = \la^2|\xv_1-\xv_2|=V_{q\bar q}^{(0)}$, \cf\ \eq{eII36}. This is expected, as the potential should not abruptly change with the emission of a gluon from a quark. When the two quarks coincide we have
\begin{align} \label{eII42}
V_{qgq}^{(0)}(\xv_1=\xv_2,\xv_g) = \sqrt{\frac{N}{C_F}}\,\la^2\,|\xv_1-\xv_g| = \frac{3}{2}\,\la^2\,|\xv_1-\xv_g|
\end{align}
reflecting the stronger potential between octet charges.
Adding the \order{g^2} term in $\mH_V$ gives using \eq{eII38},
\begin{align} \label{eII43}
V_{qgq}(\xv_1,\xv_g,\xv_2) = \frac{\la^2}{\sqrt{C_F}}\, d_{qgq}(\xv_1,\xv_g,\xv_2)+\halft\,\as\Big[\inv{N}\,\inv{|\xv_1-\xv_2|}-N\Big(\inv{|\xv_1-\xv_g|}+\inv{|\xv_2-\xv_g|}\Big)\Big]
\end{align}

\subsubsection{$qqq$ states \label{sIIB3}}

A baryon valence quark state is (for $N=3$) a superposition of the states
\begin{align} \label{eII44}
\ket{qqq} \equiv \epsilon_{ABC} \psi_A^{\dag}(\xv_1)\,\psi_B^{\dag}(\xv_2)\,\psi_C^{\dag}(\xv_3)\ket{0}
\end{align}
where a sum over the quark colors $A,B,C$ is understood. The generic terms $\propto \yv\cdot\zv$ in $h_V^{(0)}\ket{qqq}$ are   
\begin{align} \label{eII45}
\xv_1^2:& \hspace{1cm} \epsilon_{ABC} \psi_{A''}^{\dag}(\xv_1)\,\psi_B^{\dag}(\xv_2)\,\psi_C^{\dag}(\xv_3)\ket{0}T^a_{A''A'}T^a_{A'A} = \frac{4}{3}\ket{qqq} \nn\crt
\xv_1\cdot\xv_2:& \hspace{1cm} 2\,\epsilon_{ABC} \psi_{A'}^{\dag}(\xv_1)\,\psi_{B'}^{\dag}(\xv_2)\,\psi_C^{\dag}(\xv_3)\ket{0}T^a_{A'A}T^a_{B'B} = -\frac{4}{3}\ket{qqq}
\end{align}
Summing all contributions we get
\begin{align} \label{eII46}
h_V^{(0)}\ket{qqq} &= \frac{4}{3}\big[d_{qqq}(\xv_1,\xv_2,\xv_3)\big]^2\ket{qqq} \nn\crt
d_{qqq}(\xv_1,\xv_2,\xv_3) &\equiv \inv{\sqrt{2}}\sqrt{(\xv_1-\xv_2)^2+(\xv_2-\xv_3)^2+(\xv_3-\xv_1)^2}
\end{align}
The \order{\kappa^2} contribution to $\mH_V$ takes the same value as for the $q\bar q$ states \eq{eII36a} provided we choose ($C_F=4/3$)
\begin{align} \label{eII47}
\kappa_{qqq} = \frac{\la^2}{gC_F}\,\frac{1}{d_{qqq}(\xv_1,\xv_2\xv_3)}
\end{align}
The \order{g\kappa} contribution to $\mH_V$ gives the \order{\alpha_s^0} potential,
\begin{align} \label{eII48}
V_{qqq}^{(0)}(\xv_1,\xv_2,\xv_3) = g\kappa_{qqq}\,\frac{4}{3}\,\big[d_{qqq}(\xv_1,\xv_2,\xv_3)\big]^2 = \la^2 d_{qqq}(\xv_1,\xv_2,\xv_3)
\end{align}
Since $d_{qqq}(\xv_1,\xv_2=\xv_3) = |\xv_1-\xv_2|$ this $qqq$ potential reduces to the $q\bar q$ one \eq{eII36} when two of the quarks are at the same position. Adding the \order{g^2} term gives
\begin{align} \label{eII49}
V_{qqq}(\xv_1,\xv_2,\xv_3) = \la^2 d_{qqq}(\xv_1,\xv_2,\xv_3)-\frac{2}{3}\,\as\Big(\inv{|\xv_1-\xv_2|}+\inv{|\xv_2-\xv_3|}+\inv{|\xv_3-\xv_1|}\Big)
\end{align}

\subsubsection{$gg$ states \label{sIIB4}}

Finally we consider globally color singlet states of two transverse gluons,
\begin{align} \label{eII50}
\ket{g(\xv_1)g(\xv_2)} \equiv A_a^i(\xv_1)\,A_a^j(\xv_2)\ket{0}
\end{align}
Using the expression \eq{eII31} of $\mE_a(\zv)$ and the canonical commutation relation \eq{eII24} we get
\begin{align} \label{eII51}
\mE_a(\zv)\ket{g(\xv_1)g(\xv_2)} = -if_{ade}A_d^i(\xv_1)\,A_e^j(\xv_2)\ket{0}\big[\delta(\zv-\xv_1)-\delta(\zv-\xv_2)\big]
\end{align}
Operating on this with $\mE_a(\yv)$ we have
\begin{align} \label{eII52}
h_V^{(0)} \ket{g(\xv_1)g(\xv_2)} = N\,(\xv_1-\xv_2)^2\,\ket{g(\xv_1)g(\xv_2)}
\end{align}
Universality of the \order{\kappa^2} contribution to $\mH_V$ imposes
\begin{align} \label{eII53}
\kappa_{gg} = \frac{\la^2}{g\sqrt{C_F N}}\inv{|\xv_1-\xv_2|}
\end{align}
which gives
\begin{align} \label{eII54}
V_{gg}^{(0)}(\xv_1,\xv_2) = g\kappa_{gg}\,N (\xv_1-\xv_2)^2= \sqrt{\frac{N}{C_F}}\, \la^2\,|\xv_1-\xv_2|
\end{align}
As expected, this agrees with $V_{qgq}^{(0)}$ \eq{eII42} where the two quarks coincide. Adding the \order{g^2} term in $\mH_V$,
\begin{align} \label{eII55}
V_{gg}(\xv_1,\xv_2) = \sqrt{\frac{N}{C_F}}\, \la^2\,|\xv_1-\xv_2|-N\,\frac{\as}{|\xv_1-\xv_2|}
\end{align}

\section{Mesons in the rest frame \label{sIII}}

We consider $q\bar q$ states at rest bound by the \order{\alpha_s^0} linear potential $V_{q\bar q}^{(0)}$ in \eq{eII36}. For quark masses $m \alt \la$ the binding is relativistic. The \order{\as} contribution from $\ket{qg\bar q}$ Fock states with transverse gluons (which we neglect here) is then of the same order as the instantaneous gluon exchange potential $V_{q\bar q}^{(1)}$. On the other hand, for heavy quarkonia the bound state equation reduces to the non-relativistic Schr\"odinger equation. The $V_{q\bar q}^{(1)}$ potential then dominates $V_{qg\bar q}^{(1)}$, as assumed in the Cornell approach \cite{Eichten:1979ms}. 

In the temporal gauge ($A^0_a=0$) Gauss' law is implemented as a constraint rather than as an operator equation. Hence the \order{\alpha_s^0} color field does not create $q\bar q$ pairs from the vacuum. ``String breaking'' arises through a non-vanishing overlap between the meson states derived here, leading to decay and hadron loop contributions (\fig{f2}) essential for unitarity. We discuss these and related aspects in section \ref{sIV}.

\subsection{Bound state equation \label{sIIIA}}

The meson state $\ket{M}$ in \eq{eII30} defines a color reduced wave function $\Phi_{\alpha\beta}$, which is a $4\times 4$ matrix in the Dirac indices. The stationarity condition \eq{eII14} gives a bound state equation (BSE) for $\Phi$, in the \order{\alpha_s^0} approximation where $\mH_V^{(0)}$ of \eq{eII32} is the interaction term. Similarly as for Positronium (before the non-relativistic limit is taken) the meson BSE has the form \eq{eII19}, where now $V = \la^2|\xv| \equiv V'r$.

It is convenient to define
\begin{align} \label{eIII1}
&\rla_\pm \equiv \frac{2}{M-V}(i\alv\cdot\rnab+m\gz)\pm 1
&\lla_\pm \equiv (i\alv\cdot\lnab-m\gz)\frac{2}{M-V}\pm 1
\end{align}
which satisfy
\begin{align}
\rla_-\rla_+ &= \frac{4}{(M-V)^2}(-\rnab^2+m^2)-1 +\frac{4iV'}{r(M-V)^3}\,\alv\cdot\xv \,(i\alv\cdot\rnab+m\gz) \nn \crt
\lla_+\lla_- &= (-\lnab^2+m^2)\frac{4}{(M-V)^2}-1 + (i\alv\cdot\lnab-m\gz)\, \alv\cdot\xv\,\frac{4iV'}{r(M-V)^3} \label{eIII2}
\end{align}
Using this notation the bound state equation \eq{eII19} is
\begin{align} \label{eIII3}
\rla_-\Phi(\xv)+\Phi(\xv)\lla_- = 0
\end{align}
From this follows
\begin{align} \label{eIII4}
\rla_-\Phi(\xv) &= -\Phi(\xv)\lla_- = \frac{V'}{r(M-V)^2}\,\com{i\alv\cdot\xv}{\Phi(\xv)}
\end{align}
The derivation is given in appendix \ref{B}, for the general case of bound states in motion.

\subsection{Separation of radial and angular variables \label{sIIIB}}

The $4\times 4$ wave function $\Phi_{\alpha\beta}(\xv)$ may be expressed as a sum of terms with distinct Dirac structures $\Gamma_{\alpha\beta}^{(i)}(\xv)$, radial functions $F_i(r)$ and angular dependence given by the spherical harmonics $Y_{j\lambda}(\hat\xv)$:
\begin{align} \label{eIII5}
\Phi(\xv) = \sum_i \Gamma_{\alpha\beta}^{(i)}F_i(r)Y_{j\lambda}(\hat\xv)
\end{align} 
where $r=|\xv|$ and $\hat\xv=\xv/r$. Provided the Dirac structures are rotationally invariant, $\com{\Jv}{\Gamma_i(\xv)}=0$ with $\Jv=\Lv+\Sv$ \eq{A5}, the meson state will be an eigenstate \eq{A6} of the angular momentum operators of $\bs{\mJ}^2$ and $\bs{\mJ}^z$ with eigenvalues $j(j+1)$ and $\lambda$, respectively.

The $\Gamma^{(i)}(\xv)$ need contain at most one power of the Dirac vector $\alv=\gz\gv$ since higher powers may be reduced using $\alpha^i\alpha^j=\delta^{ij}+i\epsilon_{ijk}\alpha^k\gf$. Rotational invariance requires that $\alv$ be dotted into a vector. We choose as basis the three orthogonal vectors $\xv,\ \Lv=\xv\times(-i\nv)$ and $\xv\times\Lv$. Each of the four Dirac structures $1,\ \alv\cdot\xv,\ \alv\cdot\Lv$ and $\alv\cdot\xv\times\Lv$ can be multiplied by the rotationally invariant Dirac matrices $\gz$ and/or $\gf$. This gives altogether $4\times 2\times 2=16$ possible $\Gamma^{(i)}(\xv)$. Other invariants may be expressed in terms of these, \eg,
\begin{align}
i\,\alv\cdot\nv &= (\alv\cdot\xv)\,\inv{r}i\,\partial_r +\inv{r^2}\,\alv\cdot\xv\times\bs{L} \label{eIII6} \\[2mm]
(\alv\cdot\nv)(\alv\cdot\xv) &= 3+r\partial_r +\gf\,\alv\cdot\Lv \label{eIII7}
\end{align}

The $\Gamma^{(i)}(\xv)$ may be grouped according to the parity $\eta_P$ \eq{A11} and charge conjugation $\eta_C$ \eq{A15} quantum numbers that they imply for the wave function. Since $Y_{j\lambda}(-\hat\xv)=(-1)^j Y_{j\lambda}(\hat\xv)$ states of spin $j$ can belong to one of four ``trajectories'', here denoted by the parity and charge conjugation quantum numbers of their $j=0$ member\footnote{The first three trajectories were named $\pi$, $A_1$ and $\rho$ in \cite{Geffen:1977bh}.}:
\beq\label{eIII8}
\begin{array}{llcl} 0^{-+} \ \mbox{trajectory} & [s=0,\ \ell=j]: & -\eta_P=\eta_C=(-1)^{j} & \gf,\ \gz\gf,\ \gf\,\alv\cdot\xv,\ \gf\,\alv\cdot\xv\times\Lv \\[2mm]
0^{--}\ \mbox{trajectory} & [s=1,\ \ell=j]: & \eta_P= \eta_C=-(-1)^{j} & \gz\gf\,\alv\cdot\xv,\ \gz \gf\,\alv\cdot\xv\times\Lv,\ \alv\cdot\Lv,\ \gz\,\alv\cdot\Lv  \\[2mm]
0^{++}\ \mbox{trajectory} & [s=1,\ \ell=j\pm 1]: & \eta_P= \eta_C=+(-1)^{j} & 1,\ \alv\cdot\xv,\ \gz\alv\cdot\xv,\ \alv\cdot\xv\times\Lv,\ \gz\alv\cdot\xv\times\Lv,\ \gz\gf\,\alv\cdot\Lv  \\[2mm]
0^{+-} \ \mbox{trajectory} & [\mbox{exotic}]: & \eta_P=-\eta_C=(-1)^{j} & \gz,\ \gf\,\alv\cdot\Lv
\vspace{-.4cm}\end{array} 
\eeq

The non-relativistic spin $s$ and orbital angular momentum $\ell$ are indicated in brackets. Relativistic effects mix the $\ell=j\pm 1$ states on the $0^{++}$ trajectory, resulting in a pair of coupled radial equations. The $j=0$ state on the $0^{--}$ trajectory and the entire $0^{+-}$ trajectory are incompatible with the $s,\ell$ assignments and thus exotic in the quark model. They turn out to be missing also in the relativistic case. The bound state equation \eq{eIII3} has no solutions for states on the $0^{+-}$ trajectory ($\Gamma^{(i)} = \gz$ or $\gf\,\alv\cdot\Lv$) since
\begin{align} \label{eIII9}
i\nv\cdot\acom{\alv}{\gz}= i\nv\cdot\acom{\alv}{\gf\,\alv\cdot\Lv}= m\com{\gz}{\gz}=m\com{\gz}{\gf\,\alv\cdot\Lv}=0
\end{align}

\subsection{The $0^{-+}$ trajectory: $\eta_P=(-1)^{j+1}, \hspace{.3cm} \eta_C=(-1)^{j}$ \label{sIIIC}}

According to the classification \eq{eIII8} we expand the wave function $\Phi_{-+}(\xv)$ of the $0^{-+}$ trajectory states as
\begin{align} \label{eIII10}
\Phi_{-+}(\xv) = \Big[F_1(r) + i\,\alv\cdot\xv\,F_2(r) + \alv\cdot\xv\times\Lv\,F_3(r) + \gz\,F_4(r)\Big]\gf\,Y_{j\lambda}(\hat\xv)
\end{align}
Using this in the bound state equation \eq{eIII3}, noting that $i\nv\cdot \xv\times\Lv=\Lv^2$ and comparing terms with the same Dirac structure we get the conditions:
\begin{align}\label{eIII11}
\gf: \hspace{.5cm} & -(3+r\partial_r)F_2+j(j+1)F_3+m F_4=\halft (M-V)F_1 \nn \\
\gf\,\alv\cdot\xv:\hspace{.5cm} & \frac{1}{r}\partial_r F_1= \halft (M-V)F_2 \nn \\
\gf\,\alv\cdot\xv\times\Lv: \hspace{.5cm} & \inv{r^2}F_1 = \halft (M-V)F_3 \nn \\
\gz\gf: \hspace{.5cm} & m F_1 = \halft (M-V)F_4
\end{align}
Expressing $F_2,\ F_3$ and $F_4$ in terms of $F_1$ we find the radial equation (denoting $F_1' \equiv \partial_r F_1$)
\begin{align} \label{eIII12}
F_1''+\Big(\frac{2}{r}+\frac{V'}{M-V}\Big)F_1' + \Big[\quart (M-V)^2-m^2-\frac{j(j+1)}{r^2}\Big]F_1 = 0
\end{align}
in agreement with the corresponding result in Eq. (2.24) of \cite{Geffen:1977bh}.
The wave function \eq{eIII10} may be expressed as
\begin{align}\label{eIII13}
\Phi_{-+}(\xv) &= \Big[\frac{2}{M-V}(i\alv\cdot\rnab+m\gz)+1\Big]\gf\,F_1(r)Y_{j\lambda}(\hat\xv) = F_1(r)Y_{j\lambda}(\hat\xv)\,\gf\Big[(i\alv\cdot\lnab-m\gz)\frac{2}{M-V}+1\Big] \nn \crt
&= \rla_+\gf\,F_1(r)Y_{j\lambda}(\hat\xv) = F_1(r)Y_{j\lambda}(\hat\xv)\,\gf \lla_+
\end{align}
Given \eq{eIII13} we may check that the radial equation \eq{eIII12} follows using the identities \eq{eIII2} in the bound state equation \eq{eIII3}. Both the quark and antiquark contributions have a spin-dependent ($\Sv=\halft \gf\alv$) interaction which cancels in their sum. The contribution from the quark term is, taking into account the radial equation,
\begin{align} \label{eIII14}
\rla_-\Phi_{-+}(\xv) = \frac{8V'}{r(M-V)^3}\,\Sv\cdot(\rorb\,\gf-im\,\xv\,\gz) \,F_1(r)Y_{j\lambda}(\hat\xv)
\end{align}
 
\subsubsection*{Non-relativistic limit of the $0^{-+}$ trajectory wave functions \label{sIIIC1}}

The non-relativistic (NR) limit is in the rest frame defined by 
\begin{align} \label{eIII15}
\frac{V}{m} \to 0 \hspace{2cm} \frac{\partial}{\partial r} \sim \inv{r} \sim \sqrt{m\,V}
\end{align}
The binding energy $E_b \sim V$ is defined by $M = 2m + E_b$.

In the radial equation \eq{eIII12} we have
\begin{align} \label{eIII16}
\frac{V'}{M-V} = \frac{V}{r(M-V)} \ll \inv{r} \hspace{2cm} \quart (M-V)^2-m^2 \simeq m(E_b-V)
\end{align}
so in the NR limit
\begin{align} \label{eIII17}
F_{1,NR}''+\frac{2}{r}F_{1,NR}' + \Big[m(E_b-V)-\frac{j(j+1)}{r^2}\Big]F_{1,NR} = 0
\end{align}
In the wave function \eq{eIII13} we have at leading order
\begin{align} \label{eIII18}
\Lambda_+ = \frac{2}{M-V}(i\alv\cdot\nv+m\gz) +1 \simeq 1+\gz
\end{align}
giving
\begin{align} \label{eIII19}
\Phi_{-+}^{NR} = (1+\gz)\gf\,F_{1,NR}(r) \sph(\Omega)
\end{align}

\subsection{The $0^{--}$ trajectory: $\eta_P=(-1)^{j+1}, \hspace{.3cm} \eta_C=(-1)^{j+1}$ \label{sIIID}}

According to the classification \eq{eIII8} we expand the wave function $\Phi_{--}(\xv)$ of the $0^{--}$ trajectory states as
\begin{align} \label{eIII20}
\Phi_{--}(\xv) = \Big[\gz\,\alv\cdot\Lv\, G_1(r) + i\,\gz\gf\,\alv\cdot\xv\,G_2(r) + \gz\gf\,\alv\cdot\xv\times\Lv\,G_3(r) + m\alv\cdot\Lv\,G_4(r)\Big]Y_{j\lambda}(\hat\xv)
\end{align}
Collecting terms with distinct Dirac structures in the bound state equation \eq{eIII3},
\begin{align}\label{eIII21}
\gz\,\alv\cdot\Lv: \hspace{.5cm} & G_2 -(2+r\partial_r)G_3+m^2G_4=\halft (M-V)G_1 \nn \\
\gz\gf\,\alv\cdot\xv:\hspace{.5cm} & \frac{j(j+1)}{r^2} G_1= \halft (M-V)G_2 \nn \\
\gz\gf\,\alv\cdot\xv\times\Lv: \hspace{.5cm} & \inv{r^2}(1+r\partial_r)G_1 = \halft (M-V)G_3 \nn \\
m\,\alv\cdot\Lv: \hspace{.5cm} & G_1 = \halft (M-V)G_4
\end{align}
Expressing $G_2,\ G_3$ and $G_4$ in terms of $G_1$ we find the radial equation for the $0^{--}$ trajectory,
\begin{align} \label{eIII22}
G_1''+\Big(\frac{2}{r}+\frac{V'}{M-V}\Big)G_1' + \Big[\quart (M-V)^2-m^2-\frac{j(j+1)}{r^2}+\frac{V'}{r(M-V)}\Big]G_1 = 0
\end{align}
in agreement with the corresponding result in Eq. (2.38) of \cite{Geffen:1977bh}. The $0^{--}$ radial equation differs from the $0^{-+}$ one \eq{eIII12} only by the term $\propto \,V'/r(M-V)$. Using
\begin{align} \label{eIII23}
i\alv\cdot\nv\,\gv\cdot\Lv = \gz\gf\,\alv\cdot\xv\,\frac{i\Lv^2}{r^2}+\gz\gf\,\alv\cdot\xv\times\Lv \inv{r^2}(1+r\partial_r)
\end{align}
allows the wave function to be expressed in terms of the projector $\la_+$ of \eq{eIII1} as,
\begin{align}\label{eIII24}
\Phi_{--}(\xv) &=\rla_+\,\gv\cdot\rorb\,G_1(r)\,Y_{j\lambda}(\hat\xv)
=G_1(r)\,Y_{j\lambda}(\hat\xv)\,\gv\cdot\lorb\,\lla_+
\end{align}
where ${\overset{\lar}{L}}\strut^i = -i{\overset{\lar}{\partial}}\strut_k x^j\veps_{ijk}$.
The $j=0$ state on the $0^{--}$ trajectory is missing since $\Lv\,Y_{00}(\hat\xv)=0$.
The quark contribution to the bound state equation \eq{eIII3} is, with $\bs{S}=\halft \gf\alv$,
\begin{align} \label{eIII25}
\rla_-\Phi_{--}(\xv) = \frac{4V'}{r(M-V)^3}\big[\rorb^2\gz\gf-2m\,\bs{S}\cdot\xv\times\rorb\big]\, G_1(r)\,Y_{j\lambda}(\hat\xv)
\end{align}

\subsubsection*{Non-relativistic limit of the $0^{--}$ trajectory wave functions \label{sIIID1}}

The NR limit of the radial equation \eq{eIII22} reduces as in the $0^{-+}$ case to
\begin{align} \label{eIII26}
G_{1,NR}''+\frac{2}{r}G_{1,NR}' + \Big[m(E_b-V)-\frac{j(j+1)}{r^2}\Big]G_{1,NR} = 0
\end{align}
The equality of the $0^{-+}$ and $0^{--}$ eigenvalues reflects the spin $s$ independence of the NR limit, since $\ell=j$ for both. The wave function is
\begin{align} \label{eIII27}
\Phi_{--}^{NR} = (1+\gz)\alv\cdot\Lv\,G_{1,NR}(r) \sph(\Omega)
\end{align}

\subsection{The $0^{++}$ trajectory: $\eta_P=(-1)^{j}, \hspace{.3cm} \eta_C=(-1)^{j}$ \label{sIIIE}} 

According to the classification \eq{eIII8} we expand the wave function $\Phi_{++}(\xv)$ of the $0^{++}$ trajectory states in terms of six Dirac structures\footnote{The radial functions $F_i$ and $G_i$ are unrelated to those in sections \ref{sIIIC} and \ref{sIIID}.},
\begin{align} \label{eIII28}
\Phi_{++}(\xv) = \left\{ \Big[F_1(r) + i\,\alv\cdot\xv\,F_2(r) + \alv\cdot\xv\times\Lv\,F_3(r)\Big] + \gz\Big[\gf\,\alv\cdot\Lv\,G_1(r)+ i\,\alv\cdot\xv\,G_2(r)+\alv\cdot\xv\times\Lv\,G_3(r)\Big]\right\}Y_{j\lambda}(\hat\xv)
\end{align}
Collecting terms with distinct Dirac structures in the bound state equation \eq{eIII3},
\begin{align} \label{eIII29}
1: \hspace{.5cm} & -(3+r\partial_r)F_2+j(j+1)F_3 = \halft (M-V) F_1 \nn \\
\alv\cdot\xv: \hspace{.5cm} & \frac{1}{r}\partial_r F_1+mG_2 = \halft (M-V) F_2 \nn \\
\alv\cdot\xv\times\Lv: \hspace{.5cm} & \inv{r^2} F_1+mG_3 = \halft (M-V) F_3 \nn \\
\gz\gf\,\alv\cdot\Lv: \hspace{.5cm} & G_2 -(2+r\partial_r)G_3=\halft (M-V)G_1 \nn \\
\gz\,\alv\cdot\xv:\hspace{.5cm} & \frac{1}{r^2}\,j(j+1) G_1 +mF_2= \halft (M-V)G_2 \nn \\
\gz\,\alv\cdot\xv\times\Lv: \hspace{.5cm} & \inv{r^2}(1+r\partial_r) G_1 +mF_3 = \halft (M-V)G_3
\end{align}
It turns out to be convenient to express the above radial functions in terms of two new ones, $H_1(r)$ and $H_2(r)$:
\begin{align} \label{eIII30}
F_1 & =-\frac{2}{(M-V)^2}\big[\quart(M-V)^2-m^2\big]H_1-\frac{4m}{M-V}\partial_r(rH_2) \nn \\
F_2 &= -\frac{1}{r(M-V)}\partial_r H_1 + 2m H_2 \nn \\
F_3 &= -\inv{r^2(M-V)}H_1 \nn \\
G_1 &= 2 H_2 \nn \\
G_2 &= \frac{2}{r}\partial_r\Big[-\frac{m}{(M-V)^2}H_1+\frac{2}{M-V}\partial_r(rH_2)\Big]+(M-V)H_2 \nn \\
G_3 &= \frac{2}{r^2}\Big[-\frac{m}{(M-V)^2}H_1+\frac{2}{M-V}\partial_r(rH_2)\Big]
\end{align}
The bound state conditions \eq{eIII29} are satisfied provided $H_{1,2}$ satisfy the coupled radial equations,
\begin{align}
H_1''+\Big(\frac{2}{r}+\frac{V'}{M-V}\Big)H_1' + \Big[\quart (M-V)^2-m^2-\frac{j(j+1)}{r^2}\Big]H_1 &= 4m(M-V)H_2 \label{eIII31} \crt
H_2''+\Big(\frac{2}{r}+\frac{V'}{M-V}\Big)H_2' + \Big[\quart (M-V)^2-m^2-\frac{j(j+1)}{r^2}+\frac{V'}{r(M-V)}\Big]H_2 &= \frac{mV'}{r(M-V)^2}H_1 \label{eIII32}
\end{align}
These agree with Eqs. (2.48) and (2.49) for $F_2^{GS}$ and $G_1^{GS}$ of \cite{Geffen:1977bh}, when $H_1 = (M-V)F_2^{GS}$ and $H_2=-i\,G_1^{GS}/(M-V)$.

The wave function $\Phi_{++}(\xv)$ \eq{eIII28} can be expressed in terms of the $H_{1,2}(r)$ radial functions  and the $\la_+$ operators \eq{eIII1} as
\begin{align} \label{eIII33}
\Phi_{++}(\xv) &= 
\rla_+\big[-\halft H_1+2\,\gv\cdot\rorb\,\gf H_2+2im\,\alv\cdot\xv\, H_2\big]Y_{j\lambda}(\hat\xv)
+\frac{m}{M-V}\big[\rla_+\gz H_1+8H_2\big]Y_{j\lambda}(\hat\xv)\crt
&= Y_{j\lambda}(\hat\xv)\big[-\halft H_1-2 H_2\gf\,\gv\cdot\lorb\,+2im H_2\,\alv\cdot\xv\, \big] \lla_+ -Y_{j\lambda}(\hat\xv)\big[H_1\gz \lla_+-8 H_2\big]\frac{m}{M-V} \nn
\end{align}
The quark contribution to the bound state equation \eq{eIII3} is, with $\bs{S}=\halft \gf\alv$,
\begin{align} \label{eIII34}
\rla_- \Phi_{++}(\xv) &= -\frac{4V'}{r(M-V)^3}\Big[\bs{S}\cdot\rorb+\frac{m}{M-V}\gz r\partial_r\Big]H_1(r)Y_{j\lambda}(\hat\xv)+\frac{8V'}{r(M-V)^3}\big[\rorb^2+m^2r^2\big]\gz\,H_2(r)Y_{j\lambda}
\end{align}

When $m=0$ chiral symmetry implies that $\Phi(\xv)$ and $\gf\Phi(\xv)$ define bound states with the same mass $M$, as is apparent from the bound state equation \eq{eII19}. The radial equations \eq{eIII31} and \eq{eIII32} in fact decouple and coincide with the radial equations of the $0^{-+}$ \eq{eIII12} and $0^{--}$ \eq{eIII22} trajectories, respectively. The $\Phi_{++}$ wave functions correspondingly reduce to $\gf\Phi_{-+}$ and $\gf\Phi_{--}$. We discuss the case of spontaneously broken chiral symmetry in section \ref{sVII}.

\subsubsection*{Non-relativistic limit of the $0^{++}$ trajectory wave functions \label{sIIIE1}}

The radial $0^{++}$ functions $H_1$ \eq{eIII31} and $H_2$ \eq{eIII32} remain coupled in the NR limit,
\begin{align}
H_{1,NR}''+\frac{2}{r}H_{1,NR}' + \Big[m(E_b-V)-\frac{j(j+1)}{r^2}\Big]H_{1,NR} &= 8m^2 H_{2,NR} \label{eIII35} \crt
H_{2,NR}''+\frac{2}{r}H_{2,NR}' + \Big[m(E_b-V)-\frac{j(j+1)}{r^2}\Big]H_{2,NR} &= \frac{V}{4mr^2} H_{1,NR}  \label{eIII36}
\end{align}
Since $\ell = j\pm 1$ the binding energies are not expected to be the same as on the $0^{-+}$ and $0^{--}$ trajectories, for which $\ell = j$. The lhs. of both equations scale as $1/r^2 \sim mV$, implying the ratio
\begin{align} \label{eIII37}
\frac{H_{2,NR}}{H_{1,NR}} \sim \frac{V}{m}
\end{align}
In the expression \eq{eIII33} for $\Phi_{++}$ the leading contribution $\propto H_1$ vanishes for $\la_+\simeq 1+\gz$ \eq{eIII18}. This requires to retain the \order{\sqrt{V/m}} term in $\la_+$,
\begin{align} \label{eIII38}
\Lambda_+ \simeq 1+\gz + \frac{i}{m}\alv\cdot\nv
\end{align}
Then the contribution $\sim \sqrt{V/m}\,H_1 \sim \sqrt{m/V}\,H_2$ matches the leading $H_2$ contribution $m\alv\cdot\xv\,H_2 \sim \sqrt{m/V}\,H_2$. The $2\gv\cdot\Lv\,\gf H_2$ term is subdominant, as are the \order{V/m} corrections in $\Lambda_+$. This gives
\begin{align} \label{eIII39}
\Phi_{++}^{NR} = \frac{i}{2m}(1+\gz)\big[-\alv\cdot\nv H_{1,NR}(r)+4m^2\alv\cdot\xv H_{2,NR}(r)\big]\sph(\Omega)
\end{align}

Orbital angular momentum is conserved in the NR limit, implying
\begin{align} \label{eIII40}
\comb{\Lv^2}{\Phi_{++}^{NR}} = \ell(\ell+1)\Phi_{++}^{NR} \hspace{2cm} \ell = j\pm 1
\end{align}
Using
\begin{align}
\comb{\rorb^2}{\xv} &= 2\big(-\nv\,r^2+\xv\, r\partial_r+3\xv\big) \label{eIII41} \crt
\comb{\rorb^2}{\nv} &= 2\big(\xv\,\nv^2-\nv\,r\partial_r\big)  \label{eIII42}
\end{align}
gives
\begin{align} \label{eIII43}
\comb{\Lv^2}{\Phi_{++}^{NR}} &= i(1+\gz)\,\alv\cdot\nv\,\inv{2m}\big[2r H_{1,NR}'-j(j+1)H_{1,NR}-8m^2r^2 H_{2,NR}\Big]\sph \nn\crt
&+i(1+\gz)\,\alv\cdot\xv\,2m\Big[\inv{2m}(E_b-V)H_{1,NR}+2r H_{2,NR}' +2H_{2,NR} + j(j+1)H_{2,NR}\Big]\sph
\end{align}
Comparing with the Dirac structures in \eq{eIII39} and \eq{eIII40} gives two conditions,
\begin{align}
8m^2 H_{2,NR} &= \frac{2}{r}H_{1,NR}'+\frac{1}{r^2}\big[\ell(\ell+1)-j(j+1)\big]H_{1,NR} = \frac{2}{r}H_{1,NR}'+\frac{1}{r^2}\big[\pm(2j+1)+1\big]H_{1,NR}  \label{eIII44} \crt
m(E_b-V)H_{1,NR} &= -4m^2r H_{2,NR}' + \big[\pm(2j+1)-1\big]2m^2H_{2,NR} \hspace{3cm} \mbox{for}\ \ \ell=j\pm 1  \label{eIII45}
\end{align}
Using the expression \eq{eIII44} for $8m^2 H_{2,NR}$ in the radial equation \eq{eIII35} gives the expected NR radial equation,
\begin{align} \label{eIII46}
H_{1,NR}''+ \Big[m(E_b-V)-\frac{\ell(\ell+1)}{r^2}\Big]H_{1,NR} &= 0
\end{align}
To check the self-consistency of \eq{eIII44} with \eq{eIII45} we may use \eq{eIII44} to express $H_{2,NR}$ and $H_{2,NR}'$ in terms of $H_{1,NR}, H_{1,NR}'$ and $H_{1,NR}''$ and use this in \eq{eIII45}. The result agrees with \eq{eIII46}.

Using the expression \eq{eIII44} for $H_{2,NR}$ in the wave function \eq{eIII39} we have
\begin{align} \label{eIII47}
\Phi_{++}^{NR} = -\frac{i}{2m}(1+\gz)\Big\{\alv\cdot\nv H_{1,NR}(r)-\alv\cdot\xv\Big[\inv{r}H_{1,NR}' +\frac{1}{2r^2}\big[\pm(2j+1)+1\big]H_{1,NR}\Big]\Big\}\sph
\end{align}
Separating $\nv$ into its radial and angular derivatives,
\begin{align}
\alv\cdot\nv &= (\alv\cdot\xv)\,\inv{r}\partial_r -i\inv{r^2}\,\alv\cdot\xv\times\bs{L} \label{eIII48}
\end{align}
we see that the radial derivative of $H_1$ cancels, so that the $\ell=j\pm 1$ NR wave functions are,
\begin{align} \label{eIII49}
\Phi_{++}^{NR} = \frac{i}{2mr^2}(1+\gz)\Big\{\halft\alv\cdot\xv \big[\pm(2j+1)+1\big]+i\alv\cdot\xv\times\Lv\Big\}H_{1,NR}\sph
\end{align}

\section{Properties of the meson states \label{sIV}}

In this section we discuss general properties of the $q\bar q$ meson wave functions. The qualitative features are similar for all states, and are illustrated by the $0^{-+}$ trajectory.

\subsection{Orthogonality \label{sIVA}}

The overlap of two $q\bar q$ states $\ket{M_1}$ and $\ket{M_2}$ \eq{eII30} is given by the annihilation of both quark fields,
\begin{align}\label{eIV1}
\bra{M_2}M_1\rangle = \int d\xv_1 d\xv_2\,\tr\Big[\Phi_2^\dag(\xv_1-\xv_2)\Phi_1(\xv_1-\xv_2)\Big] = \big[2\pi\delta(0)\big]^3\,\int d\xv\,\tr\Big[\Phi_2^\dag(\xv)\Phi_1(\xv)\Big]
\end{align}
The trace is over the Dirac indices and the factors $2\pi\delta(0)$ appear because both states are at rest.
Orthogonality follows in the standard way \cite{Dietrich:2012un} from the bound state equations \eq{eIII3} satisfied by $\Phi_1$ and $\Phi_2^\dag$,
\begin{align} \label{eIV2}
i\nv\cdot\acom{\alv}{\Phi_1(\xv)}+m\com{\gz}{\Phi_1(\xv)} &= \big[M_1-V(\xv)\big]\Phi_1(\xv)\nn\\
-i\nv\cdot\acom{\alv}{\Phi_2^\dag(\xv)}-m\com{\gz}{\Phi_2^\dag(\xv)} &= \big[M_2-V(\xv)\big]\Phi_2^\dag(\xv)
\end{align}
Multiplying the first equation by $\Phi_2^\dag(\xv)$ from the left and the second by $-\Phi_1(\xv)$ from the right and taking the trace of their sum gives
\begin{align} \label{eIV3}
2i\nv\cdot\tr\Big(\alv\acom{\Phi_2^\dag}{\Phi_1}\Big) = (M_1-M_2)\tr\Big(\Phi_2^\dag\Phi_1\Big)
\end{align}
Integrating both sides over $\xv$ we get (assuming the integrations over space components to commute)
\begin{align} \label{eIV4}
2i\sum_{j\neq k\neq \ell}\int dx^k dx^\ell\Big|_{x^j=-\infty}^{x^j=\infty} \tr\Big(\alpha^j\acom{\Phi_2^\dag}{\Phi_1}\Big) = (M_1-M_2)\int d\xv\,\tr\Big(\Phi_2^\dag\Phi_1\Big)
\end{align}
The lhs. vanishes (see \cite{Dietrich:2012un} for $D=1+1$ dimensions and \eq{eIV5} below), implying orthogonality in \eq{eIV1} when $M_1 \neq M_2$.

\subsection{Mass spectrum \label{sIVB}}

\subsubsection{Properties of the wave function at large separations $r$ \label{sIVB1}}

\begin{figure}[h] \centering
\includegraphics[width=.2\columnwidth]{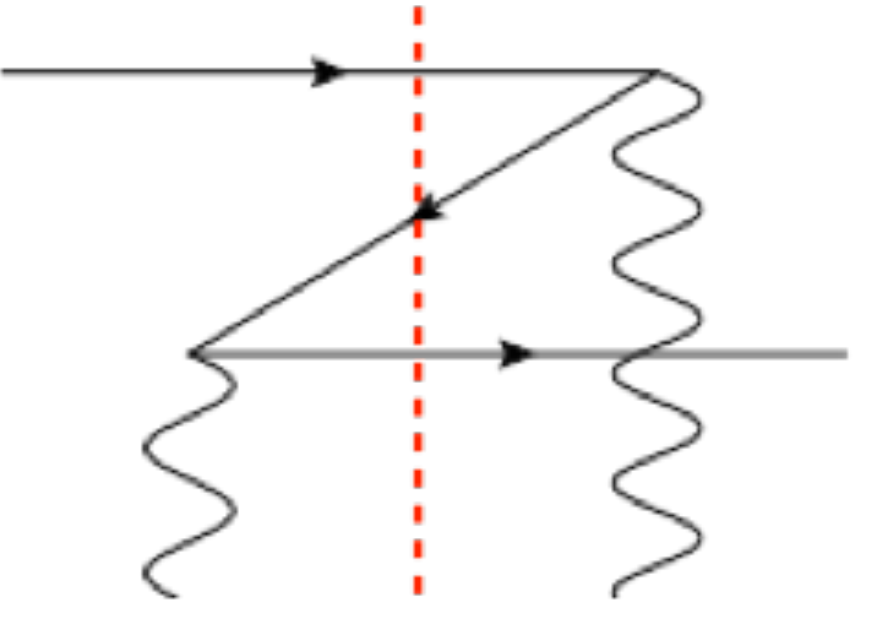}
\caption{Time-ordered ``Z''-diagram, contributing to the scattering of an electron in a strong external field. \label{f1}}
\end{figure}

Non-relativistic (Schr\"odinger) wave functions describe the probability distribution of a fixed number of bound state constituents. The consequent normalization of $\int d\xv\, |\Phi|^2$ (global norm) determines the energy eigenvalues. When the binding is relativistic ($V \agt m$) the number of constituents can change, as demonstrated by the Klein paradox \cite{Klein:1929zz} for the Dirac wave function. \fig{f1} shows how a strong external field can cause fluctuations in the instantaneous number of constituents of a bound state. For a linear potential the local norm of the Dirac electron wave function approaches a constant at large $r$ \cite{Plesset:1930zz}, reflecting the constant rate of $e^+e^-$ pair creation with increasing $V(r)$. The positrons are \textit{repelled} by the linear potential and thus are found at large $r$ (see section III of \cite{Hoyer:2016aew}). They are not confined, giving the Dirac equation a continuous energy spectrum.
 
Similarly as the Dirac wave function also the solutions of the $q\bar q$ bound state equation \eq{eIII3} have an asymptotically constant local norm. The radial wave equation \eq{eIII12} of the $0^{-+}$ trajectory determines
\begin{align} \label{eIV5}
F_1(r\to \infty) \sim \inv{r}\,r^{-im^2/V'}\exp\big[i(M-V)^2/4V'\big] \ \ \mbox{and} \ \ c.c. 
\end{align} 
Consequently the integrand (local norm) of the normalizing integral 
\begin{align}\label{eIV6}
\int d\xv\,\tr\Big[\Phi_{-+}^\dag(\xv)\Phi_{-+}(\xv)\Big] = 8\int_0^\infty dr\,r^2F_1^*(r)\Big[1-\frac{2V'}{(M-V)^3}\partial_r\Big]F_1(r)
\end{align}
tends to a constant at large $r$. This feature is common to states of all quantum numbers. The probability density similarly tends to a constant also in lower spatial dimensions ($D=1+1$ and $D=2+1$). The local norm reflects pairs created as in \fig{f1}, which have the characteristics of sea quarks. They are not constituents in the non-relativistic sense, and thus do not introduce new degrees of freedom affecting the quantum numbers of the bound state. Nevertheless, they do give rise to an increase in the parton density at small $x_{bj}$, as shown in \cite{Dietrich:2012un}.

The $q\bar q$ bound state $\ket{M}$ in \eq{eII30} generally has $b^\dag d^\dag,\ b^\dag b,\ d d^\dag$ and $d b$ operator contributions. The free $b$ and $d$ operators do not annihilate the ground state because the operators which diagonalize the Hamiltonian with an electric field are related to the free ones by a Bogoliubov transformation. This is explicitly seen in the Dirac case \cite{Hoyer:2016aew}, and is due to the $Z$-diagrams. For a linear potential the Dirac states have only positrons at large $r$. We now consider the structure of the $q\bar q$ states in the $r\to\infty$ limit.

The derivative $\nv$ in the expression \eq{eIII13} for the wave function $\Phi_{-+}(\xv_1-\xv_2)$ is equivalent to $\partial/\partial\xv_1$. After a partial integration in the state \eq{eII30} the derivative acts on $\bar\psi(\xv_1)$ (the contribution from $\nv_1 (M-V)^{-1}$ can be neglected for large $r$). If $\bar\psi_{b^\dag}\ (\bar\psi_d)$ denotes the $b^\dag\ (d)$ contribution in $\bar\psi$ we have
\begin{align}\label{eIV7}
\left.\begin{array}{c} \bar\psi_{b^\dag}(\xv_1) \crt \bar\psi_d(\xv_1) \end{array} \right\}(i\alv\cdot\lnab_1+m\gz)= \left\{\begin{array}{c} \bar\psi_{b^\dag}(\xv_1)\sqrt{-\nv_1^2+m^2} \crt -\bar\psi_d(\xv_1)\sqrt{-\nv_1^2+m^2} \end{array} \right.
\end{align}
The asymptotic behavior \eq{eIV5} implies at leading order for $r\to\infty$,
\begin{align}\label{eIV8}
\sqrt{-\nv_1^2+m^2}\,F_1(r) \simeq \halft V\,F_1(r)
\end{align}
Consequently the bracket in the wave function $\Phi_{-+}$ \eq{eIII13} becomes
\begin{align}\label{eIV9}
\Big[\frac{2}{M-V}(i\alv\cdot\rnab+m\gz)+1\Big] \simeq \left\{\begin{array}{c} -1+1=0 \crt +1+1=2 \end{array} \right. \ \ {\rm for}\ \ 
\begin{array}{c} \bar\psi_{b^\dag}(\xv_1) \crt \bar\psi_d(\xv_1) \end{array} 
\end{align}
Thus only the $d$ operator in $\bar\psi(\xv_1)$ contributes in the $r\to\infty$ limit. Similarly it can be seen that only the $b$ operator in $\psi(\xv_2)$ contributes to the state \eq{eII30}. The dominant $bd$ contribution reflects the virtual $q\bar q$ sea.

\begin{figure}[h] \centering
\includegraphics[width=.6\columnwidth]{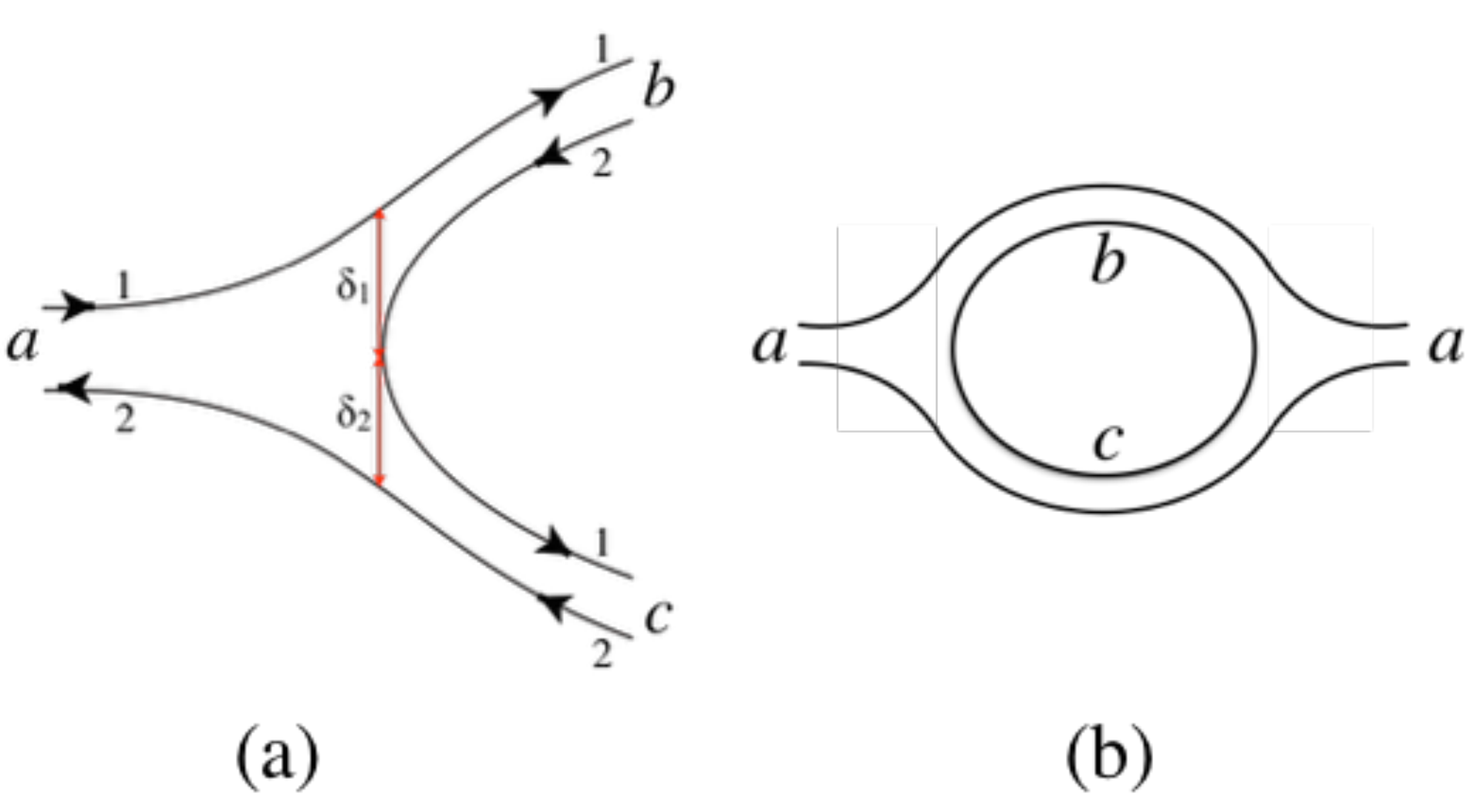}
\caption{(a) Diagram describing the meson decay $a \to b+c$, determined by the overlap of those states. Qualitatively, this may be viewed as ``string breaking'', where the color field of meson $a$ creates a $q\bar q$ pair and splits into the color fields of $b$ and $c$. \\ (b) Squaring the decay diagram gives a hadron loop correction to meson $a$. The loop will have an imaginary part when $a \to b+c$ is kinematically allowed. \label{f2}}
\end{figure}

In the present approximation the bound states are stable and have infinite radii. At large masses $M$ and/or large potentials $V(r)$ the decay and hadron loop corrections shown in \fig{f2} become important. They are  determined by the overlap of the meson states $a,\, b$ and $c$, and are essential for unitarity at the level of hadrons. Such corrections may be formulated as an expansion in $1/N_C$, with $N_C$ the number of colors. A sufficient convergence of this expansion (for $N_C=3$) is needed for the lowest approximation discussed here to be useful.
 
\subsubsection{Discrete mass spectrum \label{sIVB2}}

There can be no global normalization condition as for non-relativistic wave functions since the integral \eq{eIV6} diverges. For Dirac wave functions this means that the spectrum is continuous. Solutions of the $q\bar q$ bound state equation on the other hand are generally singular at $M-V(r)=0$, as indicated by the coefficients $\propto 1/(M-V)$ in the radial equations. A probabilistic interpretation of the wave function requires that the local norm (\ie, integrand in \eq{eIV6}) is finite for all $r$. This is the case only for discrete values of the bound state mass $M$. 

The radial equation \eq{eIII12} of the $0^{-+}$ trajectory allows $F_1(r) \sim (M-V)^\gamma$ with $\gamma=0$ and $\gamma=2$ as $M-V(r) \to 0$. The integrand in \eq{eIV6} is finite at $M-V=0$ only if $\gamma=2$. For $r \to 0$ we have as usual $F_1(r) \sim r^\beta$, with $\beta= j$ or $\beta= -j-1$. Only $\beta = j$ makes the integrand in \eq{eIV6} finite at $r=0$. The two constraints, at $M-V(r) = 0$ and $r=0$, determine the physical bound state mass spectrum.

In effect, the vanishing of $F_1(r)$ at $M-V(r) = 0$ replaces the condition of an exponential decrease with $r$ of non-relativistic wave functions. NR wave functions are defined only for $V \ll M$ and thus do not extend to $M-V=0$. The Dirac equation may be viewed as a limit of a two-particle equation where the mass $m_2$ of one particle tends to infinity, turning it into a static source. The point $V(r) = M$ (where $M$ includes $m_2$) recedes to $r=\infty$ as $m_2 \to \infty$. Hence there is no condition on the Dirac wave function at $M-V=0$.

\subsubsection{Mass spectrum of the $0^{-+}$ trajectory for $m=0$ \label{sIVB3}}

\begin{figure}[h] \centering
\includegraphics[width=1.\columnwidth]{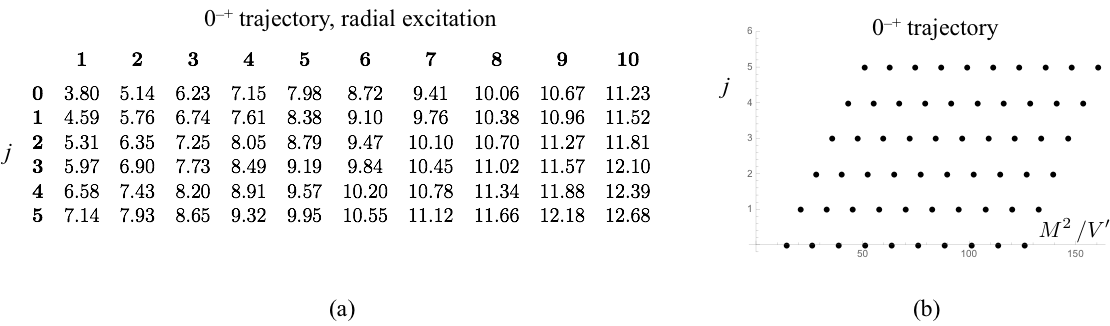}
\caption{(a) Masses $M$ of the mesons on the $0^{-+}$ trajectory for $m=0$, in units of $\sqrt{V'}$. (b) Plot of the spin $j$ {\em vs.} $M^2/V'$ for the states listed in (a). Figure taken from \cite{Hoyer:2016aew}. \label{f3}}
\end{figure}

The radial equation \eq{eIII12} can readily be solved numerically, subject to the boundary conditions $F_1(r\to 0) \sim r^j$ and $F_1(r\to M/V') \sim (M-V)^2$. As seen in \fig{f3}, for the linear potential $V(r)=V'r$ and quark mass $m=0$ the states lie on nearly linear Regge trajectories and their parallel daughter trajectories. The mass spectra of the $0^{--}$ and $0^{++}$ trajectories are similar  \cite{Hoyer:2016aew}.

\subsection{Parton picture and duality for $M \gg V(r)$ \label{sIVC}}

We expect the parton model to be applicable when the kinetic energy of a quark is large compared to its binding energy. Thus $e^+e^- \to hadrons\ $ starts (at lowest order) with the production of nearly free quarks, $e^+e^- \to q\bar q$. The subsequent hadronization process is unitary, allowing the total hadronic cross section to be calculated in terms of the initial quark production. 

According to duality $\sigma(e^+e^- \to hadrons)$ is saturated by resonances in the direct channel. This requires that the wave function of a bound state with high $M \simeq E_{CM}(e^+e^-)$ agrees with that of a free $q\bar q$ pair, at separations for which $V(r) \ll M$. It is instructive to verify this in the present approach. 

When $V(r) \ll M$ the radial equation \eq{eIII12} implies
\begin{align} \label{eIV10}
(-\nv^2+m^2)\,F_1(r)Y_{j\lambda}(\hat\xv) = \quart M^2\,F_1(r)Y_{j\lambda}(\hat\xv)
\end{align}
and thus
\begin{align} \label{eIV11}
\Big[\sqrt{-\nv^2+m^2}+i\alv\cdot\rnab+m\gz\Big]\,F_1(r)Y_{j\lambda}(\hat\xv)=\inv{M}\Big(\sqrt{-\nv^2+m^2}+i\alv\cdot\rnab+m\gz\Big)^2\,F_1(r)Y_{j\lambda}(\hat\xv)
\end{align}
In the expression \eq{eIII13} for the wave function $\Phi_{-+}(\xv)$ the derivatives operate on the $\xv=\xv_1-\xv_2$ dependence of $F_1(r)Y_{j\lambda}(\hat\xv)$. Replacing $\nv \to \nv_1$ in one of the factors on the rhs. of \eq{eIV11} and $\nv \to -\nv_2$ in the other, the $0^{-+}$ trajectory states can (after partial integrations) be expressed as
\begin{align} \label{eIV12}
\ket{M}_{V\ll M} &= \frac{2}{M^2}\int d\xv_1 d\xv_2\,\bar\psi(\xv_1)\Big[\gz\sqrt{-\nv_1^2+m^2}+i\gv\cdot\lnab_1+m\Big]\gz\gf\,F_1Y_{j\lambda}\gz\Big[\gz\sqrt{-\nv_2^2+m^2}+i\gv\cdot\rnab_2-m\Big]\psi(\xv_2)\ket{0} \nn\crt
&= \frac{2}{M^2}\int d\xv_1 d\xv_2 \int\frac{d\kv_1d\kv_2}{(2\pi)^6}\sum_{\lambda_1,\lambda_2} e^{-i(\kv_1\cdot\xv_1+\kv_2\cdot\xv_2)}\,F_1Y_{j\lambda}\big[\bar u(\kv_1,\lambda_1)\gf v(\kv_2,\lambda_2)\big] \,b_{\kv_1,\lambda_1}^\dag \,d_{\kv_2,\lambda_2}^\dag \ket{0}
\end{align}
The factors in brackets on the first line project out $b^\dag d^\dag$ from the field operators, giving a state with just the valence quark and antiquark. The $b$ and $d$ operator contributions in $\ket{M}$ are due to pair creation as in \fig{f1}, which is absent for $V \ll M$. The expression \eq{eIV12} can be further simplified using
\begin{align} \label{eIV13}
\kv_1\cdot\xv_1+\kv_2\cdot\xv_2 = \halft(\kv_1+\kv_2)(\xv_1+\xv_2) + \halft(\kv_1-\kv_2)(\xv_1-\xv_2)
\end{align}
Integrating over $\xv_1+\xv_2$ gives momentum conservation, 
\begin{align} \label{eIV14}
\ket{M}_{V\ll M} &= \frac{2}{M^2}\int d\xv \int\frac{d\kv}{(2\pi)^3}\sum_{\lambda_1,\lambda_2} e^{-i\kv\cdot\xv}\big[\bar u(\kv,\lambda_1)\gf v(-\kv,\lambda_2)\big]\,F_1(r)Y_{j\lambda}(\hat\xv) \,b_{\kv,\lambda_1}^\dag \,d_{-\kv,\lambda_2}^\dag \ket{0}
\end{align}
We may use the relation $v(-\kv,\lambda)=i\gamma^2 u^*(-\kv,\lambda)$ which is implied by charge conjugation \eq{A13} to evaluate the quark helicity dependence of the states on the $0^{-+}$ trajectory when $V \ll M$,
\begin{align} \label{eIV15}
\ket{M}_{V\ll M} &= \int d\xv \int\frac{d\kv}{(2\pi)^3}\,\frac{4E_k}{M^2} e^{-i\kv\cdot\xv}\,F_1(r)Y_{j\lambda}(\hat\xv)\sum_{\lambda_1}(-1)^{\lambda_1+1/2} \,b_{\kv,\lambda_1}^\dag \,d_{-\kv,-\lambda_1}^\dag \ket{0}
\end{align}
Expressing $E_k e^{-i\kv\cdot\xv}=\sqrt{-\nv^2+m^2}\,e^{-i\kv\cdot\xv}$ and partially integrating over $\xv$ using \eq{eIV10} gives
\begin{align} \label{eIV16}
\ket{M}_{V\ll M} &= \frac{2}{M}\int d\xv \int\frac{d\kv}{(2\pi)^3}\, e^{-i\kv\cdot\xv}\,F_1(r)Y_{j\lambda}(\hat\xv)\sum_{\lambda_1}(-1)^{\lambda_1+1/2} \,b_{\kv,\lambda_1}^\dag \,d_{-\kv,-\lambda_1}^\dag \ket{0}
\end{align}

To illuminate the structure of the state we now consider the special case of $j=m=0$. The radial wave function is then, for $V(r) \ll M$ and arbitrarily normalized,
\begin{align} \label{eIV17}
F_1(r) = \inv{r} \sin(\halft Mr) \hspace{2cm} (j=m=0)
\end{align}
The integral over $\xv$ becomes
\begin{align} \label{eIV18}
\int d\xv\,e^{-i\kv\cdot\xv}\,F_1(r) = \int_0^R dr\,r^2\,\inv{r}\sin(\halft Mr)\, \frac{4\pi}{kr}\sin(kr)
= \frac{2\pi}{k}\int_0^R dr\,\Big\{\cos\big[(\halft M-k)r\big]-\cos\big[(\halft M+k)r\big]\Big\}
\end{align}
where the range $R$ of the $r$-integration is limited by $V'R \ll M$. For $M \to \infty$ also $R\to\infty$ and the term $\cos\big[(\halft M+k)r\big]$ in the integrand is suppressed. Thus
\begin{align} \label{eIV19}
\int d\xv\,e^{-i\kv\cdot\xv}\,F_1(r) \simeq \frac{2\pi}{k}\int_0^R dr\,\cos\big[(\halft M-k)r\big] 
\simeq \frac{2\pi^2}{k}\delta(k-\halft M)
\end{align}
where the $\delta$-function is understood to limit $|k-\halft M| \lsim 1/R$. Using this in the expression \eq{eIV16} gives
\begin{align} \label{eIV20}
\ket{M}_{V\ll M} \simeq \inv{(4\pi)^{3/2}}\int d\Omega_\kv \sum_{\lambda}(-1)^{\lambda+1/2} \,b_{\kv,\lambda}^\dag \,d_{-\kv,-\lambda}^\dag \ket{0} \hspace{1cm} \mbox{where}\ \ k=\halft M
\end{align}
Thus the bound state wave function reduces to that of a free $q\bar q$ pair, isotropically distributed since we considered a $J^{PC}=0^{-+}$ state. Similarly in $e^+e^- \to hadrons$ the coupling of the virtual photon to a bound state in the direct channel will be the same as the coupling to a free $q\bar q$ pair, as required by duality.

\section{Glueballs in the rest frame  \label{sV}}

We consider states of two transversely polarized gluons $\ket{gg}$, bound by the instantaneous linear potential $V_{gg}^{(0)}$ \eq{eII54},
\begin{align}\label{eV1}
V_{gg}(r) = \sqrt{\frac{N}{C_F}}\, \la^2\,r = \frac{3}{2}\, \la^2\,r \equiv V_g'r
\end{align}
The \order{\as} instantaneous gluon exchange $V_{gg}^{(1)}$ in \eq{eII55} as well as higher Fock components ($\ket{ggg}$, $\ket{ggq\bar q}\ldots$) are ignored. Hence the Hamiltonian \eq{eII25} is approximated as $\mH=\mH_0+\mH_V$, where $\mH_V$ \eq{eII32} generates the linear potential and
\begin{align} \label{eV2}
\mH_0 &= \int d\xv\big[\halft E_{a,T}^iE_{a,T}^i +\halft A_{a,T}^i(-\delta_{ij}\nv^2+\partial_i\partial_j)A_{a,T}^j\big]
\end{align}
involves only transverse gluons $A^i_{a,T}$ and their conjugate electric fields $-E_{a,T}^i$. The canonical commutation relations \eq{eII24} imply
\begin{align} \label{eV3}
\com{\mH_0}{A^i_{a,T}(\xv)} &= iE^i_{a,T}(\xv)  &\com{\mH_0}{E^i_{a,T}(\xv)} = i\nv^2 A^i_{a,T}(\xv)
\end{align}
Consequently the bound state condition
\begin{align} \label{eV4}
(\mH_0+\mH_V)\ket{gg} = M\ket{gg}
\end{align}
requires that $\ket{gg}$ has both $A$ and $E$ components,
\begin{align} \label{eV5}
\ket{gg} &\equiv \int d\xv_1 d\xv_2 \big[A^i_{a,T}(\xv_1)A^j_{a,T}(\xv_2)\Phi^{ij}_{AA}(\xv_1-\xv_2)+A^i_{a,T}E^j_{a,T}\Phi^{ij}_{AE}+E^i_{a,T}A^j_{a,T}\Phi^{ij}_{EA}+E^i_{a,T}E^j_{a,T}\Phi^{ij}_{EE}\big]\ket{0} 
\end{align}
where sums over the color $a$ and 3-vector indices $i,j$ are understood. The constituent $A$ and $E$ fields are assumed to be normal ordered (commute with each other).

As shown in section \ref{sIIB4} the action of $\mH_V$ on $\ket{AA}$ gives the potential \eq{eV1}.
Since $\mE_a(\yv)$ \eq{eII31} has similar commutators with the $A$ and $E$ fields,
\begin{align} \label{eV6}
\com{\mE_a(\yv)}{A_d^i(\xv)} &= -i\,f_{abd}A_b^i(\xv)\delta(\xv-\yv) \nn\crt
\com{\mE_a(\yv)}{E_d^i(\xv)} &= -i\,f_{abd}E_b^i(\xv)\delta(\xv-\yv)
\end{align} 
the same potential \eq{eV1} is obtained for all four components of $\ket{gg}$ in \eq{eV5},
\begin{align} \label{eV7}
\mH_V\ket{gg} = \int d\xv_1 d\xv_2\,V_{gg}(|\xv_1-\xv_2|) \big[A_{a}(\xv_1)A_{a}(\xv_2)\Phi_{AA}(\xv_1-\xv_2) +A_{a}E_{a}\Phi_{AE} +E_{a}A_{a}\Phi_{EA} +E_{a}E_{a}\Phi_{EE}\big]\ket{0}
\end{align}
where we suppressed the 3-vector indices $i,j$ and the label $T$ of the transverse fields, which are unaffected by $\mH_0$ and $\mH_V$. Using the commutation relations \eq{eV3},
\begin{align} \label{eV8}
\mH_0\ket{gg} = i\int d\xv_1 d\xv_2\,\Big\{&\big[E_a(\xv_1)A_a(\xv_2)+A_a(\xv_1)E_a(\xv_2)\big]\Phi_{AA}(\xv_1-\xv_2)
+\big[E_aE_a+A_aA_a\nv^2\big]\Phi_{AE} \nn\crt
&+\big[A_aA_a\nv^2+E_aE_a\big]\Phi_{EA}+ \big[A_aE_a+E_aA_a\big]\nv^2\Phi_{EE}\Big\}\ket{0}
\end{align}
where $\nv$ differentiates $\Phi(\xv_1-\xv_2)$ wrt. $\xv_1-\xv_2$. 

The stationarity condition \eq{eV4} implies the following relation between the wave functions:
\begin{align} \label{eV9}
\nv^2(\Phi_{AE}+\Phi_{EA}) &= -i(M-V)\Phi_{AA} \nn\crt
\Phi_{AA}+\nv^2\Phi_{EE} &= -i(M-V)\Phi_{AE} \nn\crt
\Phi_{AA}+\nv^2\Phi_{EE} &= -i(M-V)\Phi_{EA} \nn\crt
\Phi_{AE}+\Phi_{EA} &= -i(M-V)\Phi_{EE}
\end{align}
where $V = V_{g}'|\xv_1-\xv_2|$ as in \eq{eV1}. This implies
\begin{align} \label{eV10}
\Phi_{AE} &= \Phi_{EA} = -\halft i(M-V)\Phi_{EE} \nn\crt
\Phi_{AA} &= \inv{M-V}\,\nv^2\big[(M-V)\Phi_{EE}\big] \nn\crt
\inv{M-V}\,\nv^2&\big[(M-V)\Phi_{EE}\big] + \nv^2\Phi_{EE} = -\halft (M-V)^2 \Phi_{EE}
\end{align}
If in the last equation we denote $\xv = \xv_1-\xv_2$ and $|\xv|=r$ we get
\begin{align} \label{eV11}
\nv^2\Phi_{EE}(\xv)-\frac{V_g'}{M-V}\partial_r\Phi_{EE}(\xv)-\frac{V_g'}{r(M-V)}\Phi_{EE}(\xv)+\quart (M-V)^2 \Phi_{EE}(\xv) = 0
\end{align}
Separating the radial and angular dependence according to
\begin{align} \label{eV12}
\Phi_{EE}(\xv) = F(r)Y_{\ell\lambda}(\Omega)
\end{align}
where $Y_{\ell\lambda}$ is the standard spherical harmonic function, the radial equation becomes
\begin{align} \label{eV13}
F''(r) + \Big(\frac{2}{r}-\frac{V_g'}{M-V}\Big)F'(r) +\Big[\quart(M-V)^2-\frac{V_g'}{r(M-V)}-\frac{\ell(\ell+1)}{r^2}\Big]F(r) = 0
\end{align}

There is a single dimensionful parameter $V_g'$. Scaling $r = R/\sqrt{V_g'}$ and $M = \mM\sqrt{V_g'}$ the bound state equation in terms of the dimensionless variables $R, \mM$ becomes
\begin{align} \label{eV14}
\partial_R^2 F(R) + \Big(\frac{2}{R}-\frac{1}{\mM-R}\Big)\partial_R F(R) +\Big[\quart(\mM-R)^2-\frac{1}{R(\mM-R)}-\frac{\ell(\ell+1)}{R^2}\Big]F(R) = 0
\end{align}

For $r\to 0$ we have the standard behaviors $F \sim r^\alpha$, with $\alpha = \ell$ or $\alpha = -\ell-1$. Since $\Phi_{AA} \sim \partial_r^2 \Phi_{EE}$ only the $\alpha = \ell$ solution gives a locally finite norm at $r=0$.

For $M-V\to 0$ with $F \sim (M-V)^\beta$ we have $\beta =0$, and a second solution $F \sim \log(M-V)$. Only the $\beta = 0$ solution gives a locally finite norm for $\Phi_{AA}$ at $M-V=0$.

\begin{figure}[h] \centering
\includegraphics[width=.6\columnwidth]{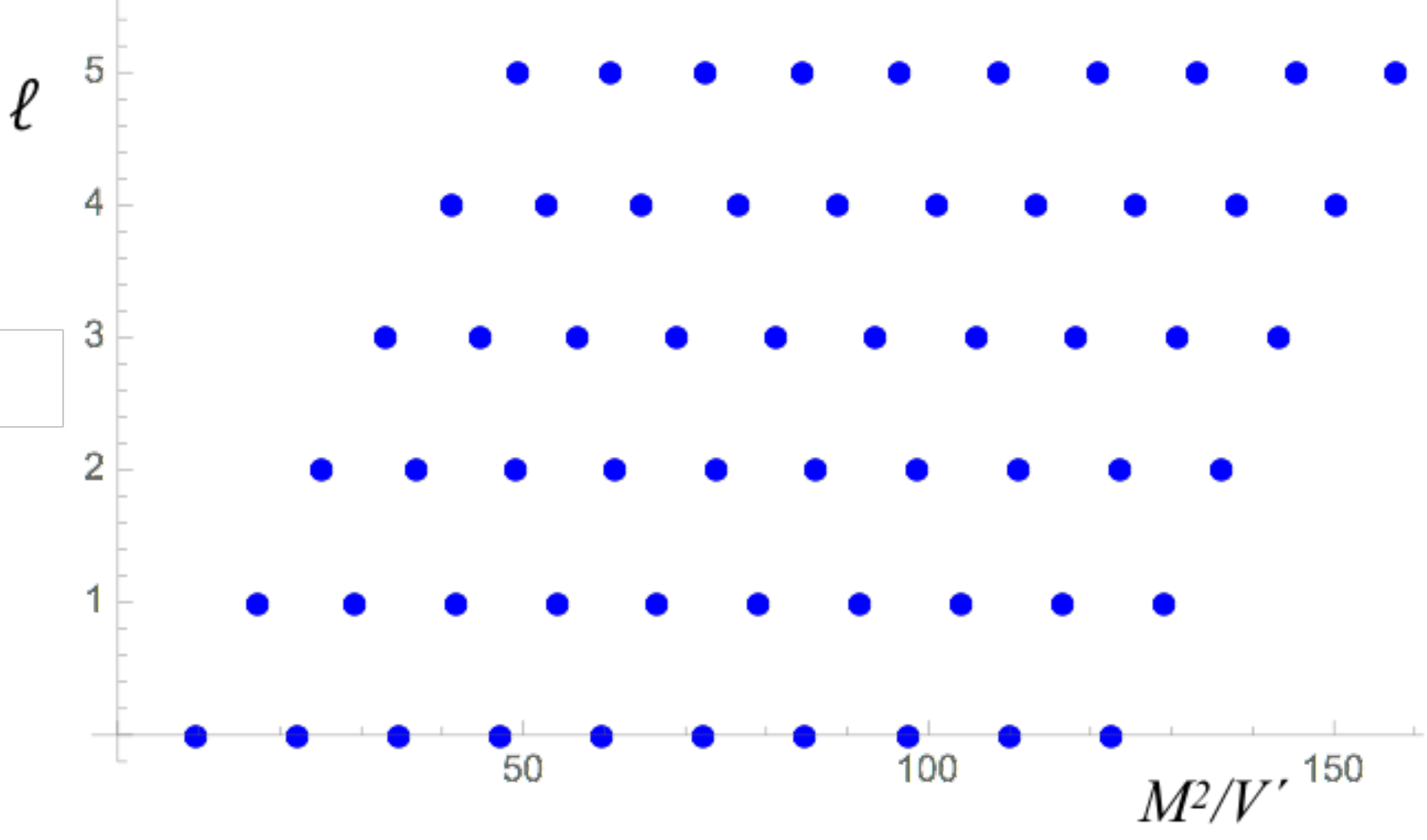}
\caption{Glueball spectrum: Orbital angular momentum $\ell$ versus $M^2/V'$. \label{f4}}
\end{figure}

The glueball states lie on approximately linear Regge and daughter trajectories (\fig{f4}). Their masses $\mM=M/\sqrt{V_g'}$ are listed in Table \ref{t1}. We may estimate the glueball masses in GeV using $\la^2 \simeq 0.18$ GeV$^2$ according to \eq{eI1}, giving $V_g' = 1.5\,\la^2 = 0.27$ GeV$^2$. Then the mass of the lowest state $M(\ell =0, n=1) = 3.10 \sqrt{V_g'} \simeq 1.6$ GeV.

 \parskip.2cm

\begin{table}[h]
$\begin{array}{c|cccccccccc}
& n=1 & n=2 & n=3 & n=4 & n=5 & n=6 & n=7 & n=8 & n=9 & n=10 \\
\hline
\ell= 0 & 3.10 & 4.70 & 5.88 & 6.87 & 7.73 & 8.50 & 9.21 & 9.870 & 10.49 & 11.07 \\
\ell= 1 &  4.14 & 5.42 & 6.46 & 7.36 & 8.16 & 8.89 & 9.57 & 10.20 & 10.80 & 11.37 \\
\ell= 2 & 5.01 & 6.08 & 7.01 & 7.84 & 8.59 & 9.28 & 9.93 & 10.54 & 11.12 & 11.67 \\
\ell= 3 & 5.75 & 6.69 & 7.54 & 8.30 & 9.01 & 9.67 & 10.29 & 10.88 & 11.43 & 11.96 \\
\ell= 4 & 6.42 & 7.26 & 8.04 & 8.76 & 9.42 & 10.05 & 10.64 & 11.21 & 11.75 & 12.26 \\
 \ell=5 & 7.02 & 7.80 & 8.52 & 9.19 & 9.82 & 10.42 & 10.99 & 11.54 & 12.06 & 12.56 \\
\end{array}$
\caption{Eigenvalues $\mM=M/\sqrt{V'}$ of the radial equation \eq{eV13}. \label{t1}}
\end{table}

\section{Mesons in motion  \label{sVI}}

\subsection{General remarks  \label{sVIA}}

Meson ($q\bar q$) states of mass $M$ and 3-momentum $\Pv$ are expressed as
\begin{align} \label{eVI1}
\ket{M,\Pv} = \inv{\sqrt{N_C}} \sum_{A,B;\alpha,\beta}\int d\xv_1 d\xv_2\,\bar\psi_\alpha^A(\xv_1)e^{i\Pv\cdot(\xv_1+\xv_2)/2}\delta^{AB}\Phi_{\alpha\beta}^{(\Pv)}(\xv_1-\xv_2)\psi_\beta^B(\xv_2)\ket{0}
\end{align}
This generalizes the $\Pv=0$ expression \eq{eII30}. The quark fields are evaluated at equal time ($t=0$) and assumed to be of the same flavor. The constraints on the wave function $\Phi^{(\Pv)}(\xv)$ following from translation, parity and charge conjugation invariance are given in appendix \ref{A}. The momentum $\Pv$ limits rotational symmetry to rotations around the direction defined by $\Pv$, which will be chosen as $z$-axis,
\begin{align} \label{eVI2}
\Pv=(0,0,P) \hspace{2cm} P=M\sinh\xi
\end{align}
All states participating in a physical process should be defined in the same frame, regardless of their momentum. The Poincar\'e invariance of a process is verified by transforming all states together to a new frame\footnote{This corresponds to an ``active'' transformation. A ``passive'' boost would describe how a state at rest appears to an observer in motion (see appendix C in v1 of this article \cite{Hoyer:2018hdj}). Active and passive transformations are distinct for interacting states.}.

We need to determine how the wave function $\Phi^{(\Pv)}(\xv)$ depends on $\Pv$. A necessary condition is that the energy of the state has the correct $\Pv$-dependence,
\begin{align} \label{eVI3}
\mH\ket{M,\Pv} = E\ket{M,\Pv} \ \ \ \text{with}\ \ \ E=\sqrt{M^2+\Pv^2}
\end{align}
This is ensured by transforming (boosting) the resting state as,
\begin{align} \label{eVI4}
\ket{M,\Pv} &= \exp(-i\xi\bs{\mK})\ket{M,0}
\end{align}
where the boost generator $\mK$ satisfies the Lie algebra relations 
\begin{align} \label{eVI5}
 \com{\mH}{\bs{\mP}}= 0 \  \hspace{2cm}
 \com{\mH}{\bs{\mK}} = i \bs{\mP}  \  \hspace{2cm}
 \com{\mP^i}{\mK^j} = i \delta^{ij}\mH 
\end{align}
For a complete verification of Poincar\'e symmetry one needs to demonstrate that the full Lie algebra is satisfied.

There are few studies of Poincar\'e transformations for equal-time bound states, even in QED. Fock states with a transversely polarized photon contribute to Positronium binding energies at leading order in $\alpha$ when $\Pv\neq 0$ \cite{Jarvinen:2004pi}. The electron-photon coupling is $\propto e \pv/m_e$, where $\pv$ is the electron momentum and $m_e$ its mass. In the rest frame $|\pv|$ is of \order{\alpha m_e}, whereas for Positronium in motion $\pv \simeq \halft\Pv$ is unsuppressed.

In $D=1+1$ dimensions the boost generator satisfying the Lie algebra \eq{eVI5} was constructed, and the wave function $\Phi^{(P)}(x)$ explicitly determined using \eq{eVI4} \cite{Dietrich:2012iy}. The same wave function was found by solving the eigenvalue condition \eq{eVI3}, requiring the local norm to be finite as in the rest frame (section \ref{sIVB2}). Both results were obtained only for a linear potential. There are no transverse photons in $D=1+1$, and the Coulomb photon exchange potential is linear. The wave function Lorentz contracts in the standard way only in the weak coupling limit ($V \ll m$). In fact, the $Z$-diagram (sea quark) contributions of \fig{f1} appear at separations $r$ which \textit{increase} with $P$ \cite{Dietrich:2012un}. The pair momenta, and thus their kinetic energy, grow with $P$. Hence their production requires a stronger field, \ie, larger $r$.

Here we find the wave function $\Phi^{(\Pv)}(\xv)$ for which the state \eq{eVI1} is an eigenstate \eq{eVI3} of the Hamiltonian in $D=3+1$, with the \order{\alpha_s^0} instantaneous potential. We have checked the result using a boost generator $\mK^z$ in \eq{eVI4} which determines the wave function in the special configuration where $\xv_1-\xv_2 \parallel \Pv$. Both results require that the \order{\alpha_s^0} potential is linear. We do not know the expression for $\mK^z$ that would satisfy the algebra \eq{eVI5} on states with general $\xv_1-\xv_2$, nor do we consider the whole Lie algebra of the Poincar\'e group. The solution for the wave function appears to be unique, making it likely that it agrees with full Poincar\'e symmetry. Including the \order{\as} gluon exchange potential requires extending \eq{eVI1} to Fock states with a transverse photon, as for Positronium  \cite{Jarvinen:2004pi}.

\subsection{Bound state equation for $\Pv \neq 0$ \label{sVIB}}

In section \ref{sIIB1} we determined the \order{\alpha_s^0} linear potential for the $\ket{q(\xv_1)\bar q(\xv_2)}$ component \eq{eII33} of a $q\bar q$ bound state. This component is characterized only by the quark positions and can be part of a bound state with any momentum. Consequently the instantaneous potential $V_{q\bar q}(\xv_1-\xv_2)$ in \eq{eII36} is independent of $\Pv$. The derivatives in the contribution of the free fermion Hamiltonian \eq{eII14a} now operate also on the plane wave exponential in \eq{eVI1}. This gives an extra term in the bound state equation implied by \eq{eVI3},
\begin{align} \label{eVI6}
i\nv\cdot\acomb{\alv}{\wfp(\xv)}-\halft \Pv\cdot \comb{\alv}{\wfp(\xv)}+m\comb{\gz}{\wfp(\xv)} &= \big[E-V(\xv)\big]\wfp(\xv)
\end{align}
where $V(\xv) = \la^2|\xv|= V'|\xv|$. 
As shown in appendix \ref{B} this BSE is equivalent to the two coupled equations
\begin{align}
\Big[\frac{2}{E-V}\big(i\alv\cdot\nv+m\gz-\halft\alv\cdot\Pv\big)-1\Big]\wfp &= -\frac{2i}{(E-V)^2}\Pv\cdot\nv\wfp+\frac{V'}{r(E-V)^2}\com{i\alv\cdot\xv}{\wfp}  \nn \crt
\wfp\Big[\big(i\alv\cdot\lnab-m\gz+\halft\alv\cdot\Pv\big)\frac{2}{E-V}-1\Big] &= \frac{2i}{(E-V)^2}\Pv\cdot\nv\wfp-\frac{V'}{r(E-V)^2}\com{i\alv\cdot\xv}{\wfp}  \label{eVI7}
\end{align}

In section \ref{sIVB2} we showed that the $\Pv=0$ wave function $\wfr$ is regular at $M=V(r)$ only for discrete (physical) values of the bound state mass $M$. At finite $\Pv$ the radial and angular dependence of the wave function cannot be separated kinematically. As we shall see, the $P$-dependence of $\wfp(\xv)$ can be found analytically on the $z$-axis, with $\Pv$ defined as in \eq{eVI2}. For $\xv=(0,0,z)$ the singular points occur at $(E-V)^2=P^2$ \cite{Hoyer:1985tz,Hoyer:1986ei}. Non-singular wave functions are obtained only when $E=\sqrt{M^2+\Pv^2}$\,.

\begin{figure}[h] \centering
\includegraphics[width=.4\columnwidth]{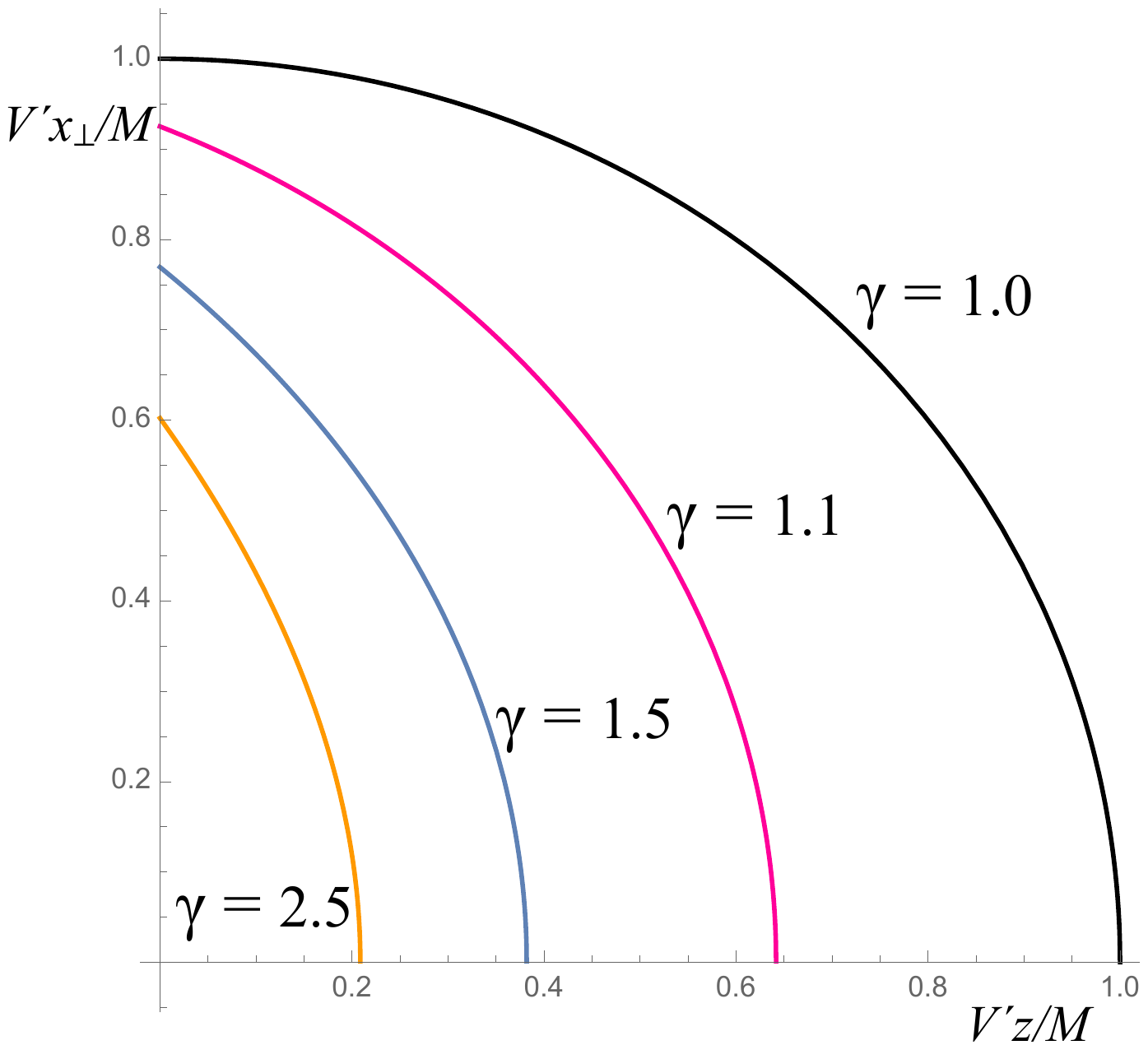}
\caption{The path $x^\perp = f(z)$ \eq{eVI11} of singularities, for various values of $\gamma = E/M$. \label{f5}}
\end{figure}

We can determine the path of singularities in the transverse plane by assuming that the most singular contribution is of power $-n$ and occurs at $x^\perp=|\xtr|=f(z)$,
\begin{align} \label{eVI8}
\wfp(\xv) = \frac{R_P(z)}{[x^\perp-f(z)]^n} + \morder{[x^\perp-f(z)]^{-n+1}}
\end{align}
where the residue $R_P(z)$ is regular. For $\wfp(\xv)$ to satisfy the first of \eq{eVI7} the terms of \order{[x^\perp-f(z)]^{-n-1}} which arise from the derivatives of the BSE acting on the denominator of \eq{eVI8} must vanish,
\begin{align} \label{eVI9}
\Big[f'(z)\Big(\alz+\frac{P}{E-V}\Big)-\inv{x^\perp}\atr\cdot\xtr\Big]R_P(z) = 0
\end{align}
Multiplying by $f'(z)\big(\alz-\frac{P}{E-V}\big)-\inv{x^\perp}\atr\cdot\xtr$ from the left gives
\begin{align} \label{eVI10}
\Big[f'^2\Big(1-\frac{P^2}{(E-V)^2}\Big)+1\Big]R_P(z) = 0
\end{align}
Hence a path $f(z)$ of singularities ($R_P \neq 0$) must satisfy 
\begin{align} \label{eVI11}
\frac{df(z)}{dz} = \pm \frac{E-V}{\sqrt{P^2-(E-V)^2}}
\end{align}
As noted above, the singular path crosses the $z$-axis ($x^\perp=0$) where $(E-V)^2=P^2$. According to \eq{eVI11} it is orthogonal to the $z$-axis at this point. A numerical solution  for $f(z)$ with $E=\sqrt{M^2+P^2}$ is shown in \fig{f5} for various values of $\gamma \equiv E/M$, in terms of the dimensionless coordinates $V'\xv/M$.

 \parskip.2cm

The BSE determines the residue function $R_P(z)$ in \eq{eVI8} in terms of a first-order differential equation. For the physical solution (see below) both $R_P(z)$ and its first derivative vanish at $x^\perp=0$. This implies $R_P(z)=0$ for the whole path.

\subsection{$\Pv$-dependence of free $q\bar q$ states \label{sVIC}}

It is instructive to consider the $P$-dependence of a free $q\bar q$ pair, \ie, $V=0$. This is trivial in the sense that it corresponds to two free Dirac states, but it illustrates how the requirement of equal time causes Lorentz contraction. The solution is relevant for the interacting case at small separations, since $V(r\to 0)=0$ for a linear potential.

Let the momenta of a free quark and antiquark be $\pv_{10} \equiv \pv_0$ and $\pv_{20} = -\pv_{0}$, respectively, in the rest frame of the pair. The total energy is then $M = 2\sqrt{\pv_0^2+m^2}$. In the frame \eq{eVI2} where the total energy is $E=M\cosh\xi$ and momentum $\Pv= M\sinh\xi\,(0,0,1)$ the quark momenta are $\pv_1 \equiv \pv$ and $\pv_2=\Pv-\pv$, with
\begin{align} \label{eVI12}
p^3 &= \halft M\sinh\xi+p_0^3\cosh\xi \nn\\
\pv^\perp &= \pv_0^\perp
\end{align} 
The state of a single quark at $t=0$ is expressed in terms of the field $\bar\psi(\xv)$ as
\begin{align} \label{eVI13}
b_{\pv,\lambda_1}^\dag\ket{0} = \int d\xv_1 \frac{d\kv_1}{(2\pi)^3 2E_1} \sum_{\mu_1}\bar u(\kv_1,\mu_1)\gz u(\pv,\lambda_1)e^{-i(\kv_1-\pv)\cdot\xv_1}b_{\kv_1,\mu_1}^\dag\ket{0}
=\int d\xv_1\,\bar\psi(\xv_1)e^{i\pv\cdot\xv_1}\gz u(\pv,\lambda_1)\ket{0}
\end{align}
With an analogous expression for the antiquark state we have
\begin{align}\label{eVI14}
\ket{M,P}_{V=0} & \equiv b_{\pv,\lambda_1}^\dag d_{\Pv-\pv,\lambda_2}^\dag \ket{0}
=\int d\xv_1 d\xv_2\,\bar \psi(\xv_1)\, e^{i\pv\cdot\xv_1+i(\Pv-\pv)\cdot\xv_2}\gz u(\pv,\lambda_1) \bar v(\Pv-\pv,\lambda_2)\gz \psi(\xv_2)\ket{0}
\end{align}
The exponent can be written
\begin{align} \label{eVI15}
\pv\cdot\xv_1+(\Pv-\pv)\cdot\xv_2 = \halft \Pv\cdot(\xv_1+\xv_2)+(\pv-\halft \Pv)\cdot(\xv_1-\xv_2)
= \halft \Pv\cdot(\xv_1+\xv_2)+\pv_0\cdot\xv_R
\end{align}
where the rest frame separation $\xv_R$ corresponding to the separation $\xv \equiv \xv_1-\xv_2$ in the moving frame was denoted
\begin{align} \label{eVI16}
\xv_R \equiv (\xv^\perp,x^3\cosh\xi)
\end{align}

The free Dirac spinors in \eq{eVI14} are related to their rest frame expressions by a boost,
\begin{align} \label{eVI17}
\gz u(\pv,\lambda_1) &= \gz \exp(\halft \xi\alz)u(\pv_0,\lambda_1) = \exp(-\halft \xi\alz)\gz u(\pv_0,\lambda_1) \nn\crt
v^\dag(\Pv-\pv,\lambda_2) &= v^\dag(-\pv_0,\lambda_2)\exp(\halft \xi\alz)
\end{align}
Using this in \eq{eVI14} the non-interacting state takes the general form \eq{eVI1}, with
\begin{align} \label{eVI18}
\wfp_{V=0}(\xv) = \exp(-\halft \xi\alz)\,\Phi_{V=0}^{(0)}(\xv_R)\exp(\halft \xi\alz)
\end{align}
Thus the boosted wave function, after extracting the factors $\exp[i\Pv\cdot(\xv_1+\xv_2)/2]$ and $\exp(\pm\halft\xi\alz)$, is given by the wave function of the rest frame at the corresponding (Lorentz dilated) quark separation \eq{eVI16}.

\subsection{$P$-dependence of $\wfp(\xtr=0,z)$ \label{sVID}}

\subsubsection{The relation \label{sVID1}}

The solution of the BSE \eq{eVI6} at $\xv^\perp=0$ can be expressed in terms of the rest frame wave function when the potential is linear \cite{Hoyer:1986ei}. This provides a boundary condition $\wfp(\xtr=0,z)$ for the partial differential equation, defining a state that is locally normalizable and has the correct $P$-dependence of the energy eigenvalue.

The $\xv^\perp=0$ solution is analogous to the one in $D=1+1$ \cite{Dietrich:2012iy,Dietrich:2012un},
\begin{align} \label{eVI19}
\exp(\halft\zeta\alz) \wfp\big[\xv^\perp=0,z_P(\tau)\big]\exp(-\halft\zeta\alz) = \Phi^{(0)}\big[\xv^\perp=0,z_0(\tau)\big]\hspace{1cm} 
\end{align}
The variable $\zeta(z)$ is defined by
\begin{align} \label{eVI20}
\cosh\zeta &= \frac{E-V(z)}{\sqrt{\Pi^2}}  &\sinh\zeta= \frac{P}{\sqrt{\Pi^2}}
\end{align}
where $E=\sqrt{M^2+P^2}$ and $V(z)=\la^2 |z|$. The kinetic 4-momentum $\Pi$ and its square are
\begin{align} \label{eVI21}
\Pi(z) &\equiv (E-V(z),\Pv)  &V'\,\tau(z) \equiv \Pi^2 = (E-V)^2-P^2= M^2-2EV+V^2
\end{align}
The variable $\tau$ is a $P$-dependent function of $z$, and takes the same value on both sides of \eq{eVI19}. In order to find $\wfp(\bs{0},z)$ at a given value of $z = z_P$ we determine the corresponding value of $\tau$ from \eq{eVI21}. Using this same value of $\tau$ we find the value $z_0$ in $\wfr(\bs{0},z_0)$ by inverting the function $\tau(z)$, now using the rest frame kinematics $E=M,\ P=0$,
\begin{align} \label{eVI22}
z_0(\tau) = (M\pm\sqrt{V'\tau})/V'
\end{align}
For $P>0$ the variable $\tau$ is negative in the range of $z$ between the singular points of the wave function, $|E-V(z)|<P$. The value of $z_0$ is then complex. This merits further study even in $D=1+1$ dimensions, where analytic solutions for the wave function are available \cite{Dietrich:2012un}. In the following we assume for simplicity that $\tau >0$.

We may check that the relation \eq{eVI19} agrees with the $P$-dependence of the free ($V=0$) case \eq{eVI18}. In \eq{eVI20} we see that $\zeta \to \xi$ as $V \to 0$, since the standard boost parameter satisfies $\cosh\xi = E/M,\ \sinh\xi=P/M$. For $V \ll M$ the variable $\tau$ is related to $z$ in the two frames according to
\begin{align} \label{eVI23}
&V'\tau(z_P) \simeq M^2-2EV'|z_P|  &V'\tau(z_0) \simeq M^2-2MV'|z_0|
\end{align}
Thus $\tau(z_P)=\tau(z_0)$ implies standard Lorentz contraction, $z_P = z_0\ M/E$ as in \eq{eVI18}.

\subsubsection{The derivation \label{sVID2}}

The BSE \eq{eVI6} may be expressed as
\begin{align} \label{eVI24}
\Big[i\rnab\cdot\alv -\halft\big(E-V+ P\alz\big)+m\gz\Big]\wfp  + \wfp \Big[i\lnab\cdot\alv -\halft\big(E-V- P\alz\big)-m\gz\Big]=0
\end{align} 
Multiplying by $\exp(\halft\zeta\alz)$ from the left and by $\exp(-\halft\zeta\alz)$ from the right the relation \eq{eVI19} requires, at $\xtr=0$,
\begin{align} \label{eVI25}
e^{\zeta\alz/2}\Big[i\rnab\cdot\alv -\halft\big(E-V+ P\alz\big)+m\gz\Big]e^{-\zeta\alz/2}\Phi^{(0)}  + \Phi^{(0)} e^{\zeta\alz/2}\Big[i\lnab\cdot\alv -\halft\big(E-V- P\alz\big)-m\gz\Big]  e^{-\zeta\alz/2}=0
\end{align}
According to \eq{eVI20} and \eq{eVI21} (for $z\equiv x^3>0$ and $\Pi^2 \geq 0$),
\begin{align} \label{eVI26}
-\halft\big(E-V\pm P\alz\big) =-\halft \Pi \exp(\pm \zeta\alz) \hspace{2cm} \Pi=\sqrt{\Pi^2}
\end{align}
Since $\nv_\perp V(|\xv|)=0$ at $\xv^\perp=0$ we may in the first term of \eq{eVI25} use
\begin{align} \label{eVI27}
i\nv_\perp\cdot\alv^\perp\,e^{-\zeta\alz/2} = e^{\zeta\alz/2}\,i\nv_\perp\cdot\alv^\perp
\hspace{1cm} \mbox{as well as} \hspace{1cm} m\gz\,e^{-\zeta\alz/2} = e^{\zeta\alz/2}m\gz
\end{align}
and analogously bring the factor $\exp(\halft\zeta\alz)$ to the right in the second term of \eq{eVI25}. 

The contribution $- i(\partial_3\,\zeta/2)\wfr$ in the first term of \eq{eVI25} cancels with the corresponding contribution from the second term. Both sides of \eq{eVI19} depend on $x^3$ only via $\tau=\Pi^2/V'$, and $\nv_\perp\, \tau=0$. Hence the relation should be $P$-independent when expressed in terms of the boost-invariant variable $\Pi$,
\begin{align} \label{eVI28}
\partial_3 = -2V'\Pi\cosh\zeta\,\partial_{\Pi^2}= -V'\cosh\zeta\,\partial_\Pi
\end{align}
The $\partial_3$ contributions in the first and second term of \eq{eVI25} are then, respectively,
\begin{align} \label{eVI29}
i\alz\rder_3\wfr &= -iV'\alz\cosh\zeta\,\rder_\Pi\wfr = -iV'e^{\zeta\alz}\alz\,\rder_\Pi\wfr +iV'\sinh\zeta\,\rder_\Pi\wfr \nn\crt
i\wfr\lder_3\alz &= -i\wfr\lder_\Pi\, \alz\cosh\zeta\, V' = -i\wfr\lder_\Pi\, \alz\,e^{-\zeta\alz}V' -iV'\sinh\zeta\,\rder_\Pi\wfr
\end{align}
The terms $\propto \sinh\zeta$ cancel. The condition \eq{eVI25} at $\xv^\perp=0$ is thus equivalent to
\begin{align} \label{eVI30}
e^{\zeta\alz}\big(-iV'\alz\rder_\Pi+i\rnab_\perp\cdot\alv^\perp-\halft\Pi+m\gz\big)\wfr
+\wfr\big(-i\lder_\Pi\,\alz V'+i\lnab_\perp\cdot\alv^\perp-\halft\Pi-m\gz\big)e^{-\zeta\alz} =0
\end{align}
Provided $V'$ is independent of $x^3$, \ie, for a linear potential, only the factors $e^{\pm\zeta\alz}$ depend explicitly on $P$. The coefficient of $\cosh\xi$ is the BSE of the rest frame, which $\wfr$ solves by definition. For \eq{eVI25} to be satisfied at all $P$ also the coefficient of $\sinh\xi$ must vanish. Given that the $\cosh\xi$ coefficient vanishes the $\sinh\xi$ condition becomes an anticommutator with $\alz$. Expressed in terms of $x^3$, which in the rest frame is related to $\Pi=M-V$,
\begin{align} \label{eVI31}
\Big\{\alz,\big[i\nv\cdot\alv+m\gz-\halft(M-V)\big]\wfr(\xv^\perp=0,x^3)\Big\} = \halft (M-V) \Big\{\alz,\rla_-\wfr(\xv^\perp=0,x^3)\Big\} = 0
\end{align}
where $\rla_-$ is defined in \eq{eIII1}. From the expressions for $\rla_-\Phi(\xv)$ in \eq{eIII14}, \eq{eIII25} and \eq{eIII34} it is clear that \eq{eVI31} holds for all wave functions at $\xtr=0$.
Thus \eq{eVI19} solves the BSE \eq{eVI6} for all $P$ at $\xv^\perp = 0$ when the potential is linear. The wave function $\wfp(\xv^\perp = 0,\tau)$ viewed as a function of $\tau$ rather then $z$ is frame independent, apart from the factors $\exp(\pm\halft\zeta\alz)$.

Since the BSE \eq{eVI6} involves derivatives of $\wfp$ the $P$-dependence in \eq{eVI19} holds also for $\nv_\perp\,\wfp$ at $\xtr=0$. Given that $\wfr$ is regular at $M-V=0$ this implies that $\wfp$ and its first derivative are regular at $E-V=\pm P$. Consequently the residue $R_P(z)$ and its derivative in \eq{eVI8} vanish at $x^\perp=0$. Since the $z$- dependence of $R_P(z)$ is given by a first-order differential equation this implies that $R_P(z)$ vanishes in the whole $\xtr$-plane, \ie, the solution specified by the boundary condition at $\xtr=0$ is locally normalizable for all $\xv$. 

We have verified that the $P$-dependence \eq{eVI19} of the wave function is consistent with a boost generator satisfying the Lie relations \eq{eVI5}, analogously as in $D=1+1$ dimensions \cite{Dietrich:2012iy}. Without the full Lie algebra there is no proof of complete Poincar\'e symmetry. However, an encouraging indication is provided by the gauge invariance of the transition electromagnetic form factor $\gamma^* a \to b$, for bound states $a$ and $b$ of any momenta. As shown in section V B of \cite{Dietrich:2012un} the matrix element of the electromagnetic current $j^\mu(z)= \bar\psi(z)\gamma^\mu \psi(z)$ (with $z$ a 4-vector),
\begin{align} \label{eVI32}
F^\mu_{ab}(z) &= \bra{M_b,\Pv_b}j^\mu(z)\ket{M_a,\Pv_a} \crt
&= e^{iz\cdot (P_b-P_a)} \int d\xv\, \Big\{e^{-i\xv\cdot(\Pv_a-\Pv_b)/2}\,\tr\big[\Phi_b^{(\Pv_b)\dag}(\xv)\gamma^\mu\gz\Phi_a^{(\Pv_a)}(\xv)\big]
-e^{i\xv\cdot(\Pv_a-\Pv_b)/2}\,\tr\big[\Phi_b^{(\Pv_b)\dag}(\xv)\Phi_a^{(\Pv_a)}(\xv)\gz\gamma^\mu\big]\Big\} \nn
\end{align}
satisfies the condition $\partial F^\mu_{ab}(z)/\partial z^\mu =0$.

\section{Spontaneous breaking of chiral invariance} \label{sVII}

We have required that the solutions of the $\Pv=0$ bound state equation \eq{eIII3} with a linear potential $V=V'|\xv|$,
\begin{align} \label{eVII1}
i\nv\cdot\acom{\alv}{\Phi(\xv)}+m\com{\gz}{\Phi(\xv)} &= \big[M-V(\xv)\big]\Phi(\xv)
\end{align}
 be locally normalizable, to allow a probabilistic interpretation of the wave function. One of the two independent radial wave functions is square integrable at $|\xv|\equiv r=0$, similarly as for the Schr\"odinger equation. A Schr\"odinger wave function determines the probability distribution of a single particle, hence its global norm $\int |\Phi(\xv)|^2 d\xv =1$. This implies a discrete mass spectrum. The relativistic equation \eq{eVII1} includes $Z$-contributions like in \fig{f1}, which increase the number of constituents. As seen in section \ref{sIVB1} the additional pairs make the local norm (integrand) tend to a constant at large $|\xv|=r$. Global normalizability in the sense of Schr\"odinger wave functions is thus neither motivated nor possible when the binding is relativistic.

Instead there arises another constraint, discussed in section \ref{sIVB2}. The wave functions (\eg, \eq{eIII13}) have factors $1/(M-V)$ which make them locally normalizable only if the radial wave function vanishes at $V(r)=V'r=M$. Together with the constraint at $r=0$ this allows only discrete bound state masses $M$. In the non-relativistic limit  locally normalizable wave functions become globally normalizable \cite{Dietrich:2012un}.

There is a special case that we did not discuss so far, namely $M=0$. Then the singular points at $r=0$ and $r=M/V'$ coincide. Locally normalizable, massless solutions exist for any quark mass $m$ \cite{Dietrich:2012un}. An $M=0$ rest frame state has vanishing four-momentum in all frames and does not correspond to a physical particle. However, the $J^{PC}=0^{++}$ ``sigma'' state may condense in the vacuum while preserving Poincar\'e invariance. This causes a spontaneous breaking of chiral invariance for massless quarks. In this section we make an exploratory study of chiral symmetry breaking with a single quark flavor (the chiral anomaly arises only at loop level). We set the scale such that $V(r) = r$, \ie,
\begin{align} \label{eVII2}
V'=1
\end{align}

\subsection{Vanishing quark mass, $m=0$} \label{sVIIA}

The states discussed in section \ref{sIII} have exact chiral symmetry for vanishing quark mass. The coupled radial equations of the $0^{++}$ trajectory \eq{eIII31} and \eq{eIII32} decouple when $m=0$, with the equation for $H_1(r)$ reducing to \eq{eIII12} for $F_1(r)$ of the $0^{-+}$ trajectory. The radial equation of the $0^{++}$ and $0^{-+}$ states with $M=m=0$,
\begin{align} \label{eVII3}
H_1''(r) + \inv{r}H_1'(r)+\inv{4}r^2 H_1(r) = 0
\end{align}
can be solved analytically. The wave functions of the $0^{++}$ ``sigma'' \eq{eIII33} and $0^{-+}$ ``pion'' \eq{eIII13} states are
\begin{align}
\inv{N_\sigma}\Phi_\sigma(\xv) &= J_0(\quart r^2)+ i\,\alv\cdot\xv\, \inv{r}J_1(\quart r^2) \hspace{2cm} (m=M=0) \label{eVII4}\crt
\inv{N_\pi}\Phi_\pi(\xv) &= \inv{N_\sigma}\,\gf\,\Phi_\sigma(\xv) \label{eVII5}
\end{align}
where $J_{0,1}$ are Bessel functions and $N_\sigma,\,N_\pi$ normalization constants. The sigma state is thus (at $t=0$, with color and Dirac indices suppressed and contractions removed),
\begin{align} \label{eVII6}
\ket{\sigma} &\equiv\hat\sigma\ket{0} = \int d\xv_1\,d\xv_2\,\bar\psi(\xv_1)\,\Phi_{\sigma}(\xv_1-\xv_2)\,\psi(\xv_2)\ket{0}  
\end{align}
Similarly $\Phi_\pi$ determines the pion state $\ket{\pi}=\hat\pi\ket{0}$.
These states have vanishing four-momentum in all frames,
\begin{align} \label{eVII7}
\hat P^\mu\ket{\sigma}=0
\end{align}

The sigma state has vacuum quantum numbers and is annihilated by the operator $\bar\psi\psi$,
\begin{align} \label{eVII8}
\bra{0}\sum_\alpha\bar\psi_\alpha(\xv)\psi_\alpha(\xv)\ket{\sigma} = \tr[\gz\Phi_\sigma(\bs{0})\gz] = 4N_\sigma
\end{align}
We define a ``chiral condensate vacuum'' in terms of $\hat\sigma$ \eq{eVII6},
\begin{align} \label{eVII9}
\ket{\chi} \equiv \hat\chi\ket{0} = \exp(\hat\sigma)\ket{0} 
\end{align}
The expectation value of $\bar\psi\psi$ when it annihilates on any one $\hat\sigma$ in $\ket{\chi}$ is,
\begin{align} \label{eVII10}
\bra{\chi}\bar\psi\psi\ket{\chi} = 4N_\sigma \bra{\chi}\chi\rangle
\end{align}
implying that chiral symmetry is spontaneously broken in the chiral condensate vacuum.
 
An infinitesimal chiral transformation $U_\chi(\beta)$ $(\beta \ll 1)$
transforms the quark fields as
\begin{align} \label{eVII11}
U_\chi(\beta)\,\bar\psi(\xv)\, U_\chi^\dag(\beta) = \bar\psi(\xv)(1-i\beta\gf) \hspace{3cm} U_\chi(\beta)\,\psi(\xv)\, U_\chi^\dag(\beta) = (1-i\beta\gf)\psi(\xv)
\end{align}
With $\Phi_\pi=\gf\Phi_\sigma = \halft\acom{\gf}{\Phi_{\sigma}}$ (absorbing a relative normalization in $\beta$) we get, since $\com{\hat\pi}{\hat\sigma}=0$,
\begin{align} \label{eVII12}
U_\chi(\beta)\ket{\chi}=\exp\Big[\int d\xv_1 d\xv_2\, \bar\psi(\xv_1)\big[\Phi_{\sigma}-i\beta\acom{\gf}{\Phi_{\sigma}}\big]\psi(\xv_2)\Big]\ket{0} = (1-2i\beta\,\hat\pi)\ket{\chi}
\end{align}
Thus a chiral transformation of $\ket{\chi}$ creates massless pions.

\subsection{Small quark mass} \label{sVIIB}

In QCD the small $u,\,d$ quark masses break chiral invariance explicitly and give the pion its physical mass. Let us consider the case $0 < m \ll 1$ (in units of $\sqrt{V'}$ \eq{eVII1}). The exact massless $(M=0)$ solution for the $0^{++}$ sigma wave function for any $m$ is\footnote{We use lowercase letters for the radial functions of the $0^{++}$ trajectory defined in \eq{eIII28}, to distinguish them from the radial functions of the $0^{-+}$ trajectory \eq{eIII10}.}
\begin{align} \label{eVII13}
\Phi_\sigma(\xv) &= f_1(r)+i\,\alv\cdot\xv\,f_2(r)+i\,\gv\cdot\xv\,g_2(r) \hspace{2cm} (M=0) \crt
\inv{N_\sigma}f_1(r) &= -\inv{r^2} e^{-ir^2/4}\Big[(2m^2-r^2)L_{(im^2-1)/2}(\halft ir^2)-2m^2L_{(im^2+1)/2}(\halft ir^2)\Big] = J_0(\quart r^2) + \morder{m^2} \nn\crt
\inv{N_\sigma}f_2(r) &= -\frac{2}{r^3} e^{-ir^2/4}\Big[(\halft ir^2-1)L_{(im^2-1)/2}(\halft ir^2)+L_{(im^2+1)/2}(\halft ir^2)\Big] = \inv{r} J_1(\quart r^2) + \morder{m^2} \nn\crt
g_2(r) &= -\frac{2m}{r}\,f_2(r) \nn
\end{align}
where the $L_\nu(x)$ are Laguerre functions. Since the sigma state remains massless when $m \neq 0$ it may form a chiral condensate as above, without breaking Poincar\'e invariance.

For the record, let us note that the massless ($M=0$) $0^{-+}$ state for finite quark mass $m$ has the wave function
\begin{align} \label{eVII14}
\Phi_\pi(\xv) &= \big[F_1(r)+i\,\alv\cdot\xv\,F_2(r)+\gz\,F_4(r)\big]\gf \crt
\inv{N_\pi}F_1(r) &= e^{-ir^2/4}L_{(im^2-1)/2}(\halft ir^2) = J_0(\quart r^2) + \morder{m^2}  \hspace{2cm} (M=0) \nn\crt
\inv{N_\pi}F_2(r) &= \frac{2}{r^3} e^{-ir^2/4}\Big[(im^2+1-\halft ir^2)L_{(im^2-1)/2}(\halft ir^2)-(im^2+1)L_{(im^2+1)/2}(\halft ir^2)\Big] = \inv{r} J_1(\quart r^2) + \morder{m^2} \nn\crt
F_4(r) &= -\frac{2m}{r}\,F_1(r) \nn
\end{align}
In the limit $r\to 0)$ $F_1(r)$ approaches a constant. Hence $F_4(r \to 0) \propto m/r$ is singular.

Consider now a $0^{-+}$ state with a non-zero mass $M$, which approaches the $M=0$ solution \eq{eVII5} in the $m \to 0$ limit. According to \eq{eIII10} also the $M \neq 0$ pion wave function has the form \eq{eVII14} and $F_1$ satisfies the radial equation \eq{eIII12},
\begin{align} \label{eVII15}
F_1''+\Big(\frac{2}{r}+\frac{1}{M-r}\Big)F_1' + \big[\quart (M-r)^2-m^2\big] F_1 = 0
\end{align}
with $F_1(r\to0) \propto r^0 \neq 0$. The constraint \eq{eIII11}, $mF_1(r) = \halft (M-r)F_4(r)$, implies a qualitative difference compared to the $M=0$ solution \eq{eVII14}: The $F_4(r)$ radial function is now finite at $r=0$,
\begin{align} \label{eVII16}
F_4(0) = \frac{2m}{M} F_1(0)
\end{align}

As a Goldstone boson the pion should be annihilated by the axial vector current $j^\mu_5(x)=\bar\psi(x)\gamma^\mu\gf\psi(x)$ and by its divergence $\partial_\mu j^\mu(x)= 2im\, \bar\psi(x)\gf\psi(x)$,
\begin{align}
\bra{\chi}\bar\psi(x)\gamma^\mu\gf\psi(x)\,\hat\pi\ket{\chi} &= iP^\mu f_\pi\,e^{-iP\cdot x} \label{eVII17} \crt
\bra{\chi}\bar\psi(x)\gf\psi(x)\,\hat\pi\ket{\chi} &= -i\,\frac{M^2}{2m}\,f_\pi\,e^{-iP\cdot x} \label{eVII18}
\end{align}
A pion of momentum $\Pv$ is an eigenstate of the Hamiltonian with eigenvalue $P^0= E= \sqrt{\Pv^2+M^2}$. The pion state at time $t$ is, using \eq{eVI1},
\begin{align} \label{eVII19}
\hat\pi(t)\ket{\chi} &= e^{-iP^0 t}\int d\xv_1 d\xv_2\,\bar\psi(t,\xv_1) e^{i\Pv\cdot(\xv_1+\xv_2)/2}\Phi_\pi^{(\bs{\Pv})}(\xv_1-\xv_2)\psi(t,\xv_2)\ket{\chi}
\end{align}
Contracting the quark fields the lhs. of \eq{eVII17} becomes
\begin{align} \label{eVII20}
\bra{\chi}\bar\psi(x)\gamma^\mu\gf\psi(x)\,\hat\pi\ket{\chi} &=
\tr\big[\gamma^\mu\gf\gz\Phi_\pi^{(\Pv)}(0)\gz\big]\,e^{-iP\cdot x}
\end{align}
According to \eq{eVI18} the momentum dependence of the wave function at the origin (where $V=0$) is 
\begin{align}  \label{eVII21}
\Phi_\pi^{(\Pv)}(\xv=0) = \exp(-\halft \bs{\xi}\cdot\alv) \Phi_\pi^{(0)}(\xv=0)\exp(\halft \bs{\xi}\cdot\alv)
\end{align}
where $\bs{\xi}$ is the boost parameter defined by $\Pv$ (\cf\ \eq{eVI2}). Both $F_1(0)$ and $F_2(0)$ are finite for the  $M>0$ solution which is normalizable at $r=0$. Thus $\alv\cdot\xv F_2(r)$ in \eq{eVII14} vanishes at $\xv=0$, and
\begin{align} \label{eVII22}
\Phi_\pi^{(0)}(0) &= \big[F_1(0)+\gz\,F_4(0)\big]\gf 
\end{align}
giving
\begin{align} \label{eVII23}
\gz\Phi_\pi^{(\Pv)}(0)\gz = -\big[F_1(0)+\slashed{P}F_4(0)/M\big]\gf
\end{align}
Substituting this into \eq{eVII20} we get
\begin{align} \label{eVII24}
\bra{\chi}\bar\psi(x)\gamma^\mu\gf\psi(x)\,\hat\pi\ket{\chi} &= \tr\big\{\gamma^\mu[-F_1(0)+\slashed{P}F_4(0)/M]\big]\big\}\,e^{-iP\cdot x} = 4P^\mu F_4(0)/M\,e^{-iP\cdot x}
\end{align}
Similarly
\begin{align} \label{eVII25}
\bra{\chi}\bar\psi(x)\gf\psi(x)\,\hat\pi\ket{\chi} &= \tr\big\{[-F_1(0)+\slashed{P}F_4(0)/M]\big]\big\}\,e^{-iP\cdot x} = -4 F_1(0)\,e^{-iP\cdot x}
\end{align}
Comparing with the rhs. of \eq{eVII17} and \eq{eVII18} we have the two conditions
\begin{align} \label{eVII26}
& F_4(0) = \quart iMf_\pi   &F_1(0) = i\,\frac{M^2}{8m}\,f_\pi 
\end{align}
which are consistent with the relation \eq{eVII16} required by the bound state equation.
A smooth $m \to 0$ chiral limit implies $M^2 \propto m$ in \eq{eVII18}. 

In section \ref{sIVB2} we saw that $F_1(V'r\to M) \propto (M-V)^\gamma$, with $\gamma = 0,\, 2$. Local normalizability at $M-V=0$ required $\gamma=2$, which together with the constraint $F_1(r\to 0) \propto r^0$ implied discrete masses $M$. Here $M$ is not fixed because we did not impose the $\gamma=2$ constraint. At small $m$ and $M$ continuity requires the same behavior at $r=0$ and $r=M/V'$, hence now $F_1(V'r\to M) \propto (M-V)^0$. On the other hand we also neglected the changes in the bound state equation arising from the chiral condensate vacuum. Further study is needed concerning effects of chiral symmetry breaking on the hadron spectrum in general, and on the Goldstone pion in particular.

\section{Summary \label{sVIII}}

We considered whether a solution to the confinement puzzle might not, after all, be found using perturbative bound state methods. This is motivated by the experimentally observed similarities of hadrons and atoms, especially for heavy quarkonia. When applicable, perturbation theory is a powerful tool. Much of our understanding of physical gauge theories is based on perturbative expansions.

We used QED as a guide (section \ref{sIIA}), as is commonly done in introducing field theory methods. The principles of bound states are often omitted in textbooks, perhaps because of the prevailing belief that hadrons are fundamentally different from atoms. The omission is unfortunate, if only because bound state perturbation theory brings qualitatively new insights to the structure of gauge theory.

Bound state methods have been developed for atoms since the beginnings of quantum mechanics. Today high-order contributions are commonly calculated in the framework of non-relativistic QED (NRQED) \cite{Caswell:1985ui,Kinoshita:1998jfa}. Our aim is not to improve on those calculations, but rather to address the principles in the choice of their starting point, the Schr\"odinger equation with the classical potential. Because the wave function is non-polynomial in $\alpha$ there is a multitude of formally equivalent bound state expansions, distinguished by their lowest approximation \cite{Caswell:1978mt,Lepage:1978hz}.

Binding energies are measurable and have a unique expansion in $\alpha$ and $\log\alpha$, which is mirrored in the Fock expansion of the bound state. For Positronium the $\ket{e^+e^-}$ Fock state suffices to determine the binding energy at lowest order in $\alpha$, whereas states such as $\ket{e^+e^-\gamma}$, $\ket{e^+e^-e^+e^-},\ldots$ contribute higher order corrections. This hierarchy of Fock states is possible in gauge theories because of the instantaneous interaction, which does not add Fock constituents.

Bound state calculations commonly use Coulomb gauge ($\nv\cdot\Av=0$) \cite{Feinberg:1977rc}. Gauss' law is then an operator equation, and the instantaneous $A^0$ field creates particles. We found temporal gauge ($A^0=0$) \cite{Strocchi:2013awa,Willemsen:1977fr,Bjorken:1979hv,Christ:1980ku,Leibbrandt:1987qv} to better reflect the Fock state hierarchy. Gauss' law then takes the form of a constraint on physical states, which serves to fix the remaining gauge degrees of freedom (time independent gauge transformations). Temporal gauge has not (to our knowledge) previously been used for bound states, so our approach should be verified by a higher order calculation.

Applications to hadrons may start by considering confinement for non-relativistic quarkonia. The neglect of heavy quark pair production simplifies the analysis. Quarkonium phenomenology \cite{Eichten:2007qx} indicates that Fock states with light quarks and gluons are suppressed (except for the higher lying $X,Y,Z$ states \cite{Chen:2016qju}, which may be hadron molecules). The suppression of higher Fock states is dynamic, since quarkonium binding energies are much larger than light quark masses. In our perturbative framework light quark and gluon Fock states are suppressed by powers of $\as$.

In this approach confinement can arise only through a homogeneous solution of Gauss' constraint. Poincar\'e invariance specifies the sourceless solution up to a universal constant $\la$ (section \ref{sIIB}). The corresponding potential is exactly linear for $q\bar q$ and $gg$ states, and confining also for states with more constituents. The growth of the potential is limited by the creation of light quark or gluon pairs (string breaking). This follows from the overlap of states (\fig{f2}), which needs to be considered also for unitarity at hadron level.

Hadrons with light quarks ($m\alt \la$) are relativistically bound by the linear potential. The coupling $\as$ is frozen at low scales and remains perturbative. The QCD Hamiltonian defines the relativistic dynamics (sections \ref{sIII} and \ref{sV}). The $\ket{q\bar q}$ mesons have quantum numbers that are compatible with the quark model (section \ref{sIIIB}). The linear potential generates virtual pairs through $Z$-diagrams as in \fig{f1}. They cause the local norm of the $q\bar q$ wave function to approach a constant at large quark separations (section \ref{sIVB1}). The pairs have the properties of sea quarks \cite{Dietrich:2012un}.

The norm $|\Phi(\xv)|^2$ of the wave function must be finite to allow a probabilistic interpretation. This imposes a condition at $M-V(|\xv|)=0$ which can be satisfied only for discrete masses $M$ (section \ref{sIVB2}).  The condition generalizes the requirement of a finite global norm $\int d\xv\,|\Phi(\xv)|^2$ for non-relativistic states.  Both the $\ket{q\bar q}$ meson (for $m=0$, \fig{f3}) and the $\ket{gg}$ glueball states (\fig{f4}) lie on approximately linear Regge trajectories and their daughters.

The instantaneous potential is determined by the positions $\xv_1,\xv_2$ of the charges and is independent of the 3-momentum $\Pv$ of the state (section \ref{sVI}). For a linear potential the $\Pv$-dependence of the meson wave functions $\wfp(\xv)$ can be expressed analytically \eq{eVI19} when the quark separation $\xv \parallel \Pv$. This provides a boundary condition for the BSE \eq{eVI6}. The norm $|\wfp(\xv)|^2$ is finite for all $\xv$ and $\Pv$ when $E=\sqrt{M^2+\Pv^2}\,$.

There are massless $(M=0)$ states with regular norm (section \ref{sVII}). They do not correspond to physical particles as $E=\Pv=0$ in all frames. However, the massless $0^{++}$ sigma state may mix with the perturbative vacuum \eq{eVII9}, maintaining Poincar\'e invariance while causing a spontaneous breaking of chiral symmetry for small quark mass $m$. The PCAC relations \eq{eVII17} and \eq{eVII18} were shown to hold, motivating further studies.

Clearly many more checks of the present approach to bound states are required, and further applications remain to be studied.


\acknowledgments

My work on these topics has relied on discussions with many colleagues, among them Jean-Paul Blaizot, Stan Brodsky, Dennis D.~Dietrich, Matti J\"arvinen, J\"orn Knoll and Stephane Peign\'e. During the preparation of this material I enjoyed visits to  ECT* (Trento), Jlab (Newport News) and CP$^3$ (Odense). I am grateful for their hospitality, and to the Department of Physics at Helsinki University for my privileges of Professor Emeritus. An annual travel grant from the Magnus Ehrnrooth Foundation has allowed me to maintain contacts and present my research to colleagues.

\appendix
\renewcommand{\theequation}{\thesection.\arabic{equation}}

\section{Symmetries of the $q\bar q$ wave function \label{A}}

In this appendix we note the transformation of $q\bar q$ (meson) states under space translations, rotations, parity and charge conjugation. The quark and antiquark are assumed to have the same flavor. The $t=0$ states with momentum $\Pv$ are described by a (color reduced) wave function $\Phi^{(\Pv)}(\xv)$ \eq{eVI1},
\begin{align} \label{A1}
\ket{M,\Pv} = \inv{\sqrt{N_C}} \sum_{A,B;\alpha,\beta}\int d\xv_1 d\xv_2\,\bar\psi_\alpha^A(\xv_1)e^{i\Pv\cdot(\xv_1+\xv_2)/2}\delta^{AB}\Phi_{\alpha\beta}^{(\Pv)}(\xv_1-\xv_2)\psi_\beta^B(\xv_2)\ket{0}
\end{align}
The states are invariant under global gauge transformations of the quark field, $\psi^B \to U^{BB'}\psi^{B'}$, with $U \neq U(\xv)$.

\subsection{Space translations \label{sA1}}

Under space translations $\xv \to \xv+\ellv$ the quark fields are transformed by the operator
\beq\label{A2}
U(\bs{\ell}) = \exp[-i\ellv\cdot\bs{\mP}] \hspace{1cm} {\rm where} \hspace{1cm} \bs{\mP} = \int d\xv\,\psi^\dag(\xv)(-i\nv)\psi(\xv)
\eeq 
The momentum operator satisfies
\beq\label{A3}
\com{\bs{\mP}}{\psi(\xv)} = i\rnab \psi(\xv)  \hspace{2cm}  \com{\bs{\mP}}{\bar\psi(\xv)} = \bar\psi(\xv)i\lnab
\eeq
With $\bs{\mP}\ket{0}=0$ we have $\bs{\mP}\ket{M,\Pv} = \Pv\ket{M,\Pv}$.

\subsection{Rotations \label{sA2}}

Rotations are generated by the angular momentum operators
\begin{align}
\bs{\mJ} &= \int d\xv\,\psi^\dag(\xv)\,\bs{J}\,\psi(\xv) \hspace{2.9cm} \bs{\mJ}^2 = \int d\xv\,\psi^\dag(\xv)\,\bs{J}^2\,\psi(\xv) \label{A4} \crt
\bs{J} &= \Lv+\Sv= \xv\times(-i\nv)+\halft\gf\gz\gv \label{A5}
\end{align}
The angular momentum quantum numbers $j,\lambda$ of a state are defined by,
\begin{align} \label{A6}
\bs{\mJ}^2\ket{j,\lambda} = j(j+1)\ket{j,\lambda} \hspace{2cm} \mJ^z\ket{j,\lambda}= \lambda\ket{j,\lambda}
\end{align}
Rest frame $(\Pv=0)$ states are invariant under rotations provided the wave function in \eq{A1} satisfies
\begin{align} \label{A7}
\com{\Jv^2}{\Phi^{(0)}(\xv)} = j(j+1)\,\Phi^{(0)}(\xv) \hspace{2cm} \com{J^z}{\Phi^{(0)}(\xv)} = \lambda\,\Phi^{(0)}(\xv)
\end{align}

\subsection{Parity \label{sA3}}

The parity operator $\bP$ reverses 3-momenta $\pv$ but leaves the spin components $\lambda$ invariant:
\beq\label{A8}
\bP b(\pv,\lambda)\bP^\dag = b(-\pv,\lambda) \hspace{2cm} \bP d(\pv,\lambda)\bP^\dag = -d(-\pv,\lambda)
\eeq
The intrinsic parity of the quarks is irrelevant for $q\bar q$ states. The {\em relative} intrinsic parity $-1$ of quarks and antiquarks in \eq{A8} ensures that the field transforms as
\beq\label{A9}
\bP \psi(t,\xv)\bP^\dag = \gz\psi(t,-\xv) \hspace{2cm} \bP \bar\psi(t,\xv)\bP^\dag = \bar\psi(t,-\xv)\gz
\eeq
Parity reverses the momentum of a state,
\beq\label{A10}
\bP\ket{M,\Pv} = \int d\xv_1\,d\xv_2\,\bar\psi(\xv_1)e^{-i\Pv\cdot(\xv_1+\xv_2)/2}\gz\Phi^{(\Pv)}(-\xv_1+ \xv_2)\gz \psi(\xv_2)\ket{0} = \eta_P \ket{M,-\Pv}
\eeq
provided the wave function satisfies
\beq\label{A11}
\gz\Phi^{(\Pv)}(-\xv)\gz = \eta_P \Phi^{(-\Pv)}(\xv) \hspace{2cm} (\eta_P = \pm 1)
\eeq
For $\Pv=0$ parity transforms the wave function into itself.

\subsection{Charge conjugation \label{sA4}}

The charge conjugation operator $\bC$ transforms particles into antiparticles,
\beq\label{A12}
\bC b(\pv,\lambda)\bC^\dag = d(\pv,\lambda) \hspace{2cm} \bC d(\pv,\lambda)\bC^\dag = b(\pv,\lambda)
\eeq 
In the Dirac representation of the $\gamma$ matrices this implies (here $T$ indicates transpose and $\aly \equiv \gz\gamma^2$)
\beq\label{A13}
\bC \psi(t,\xv)\bC^\dag = -i\aly\bar\psi^T(t,\xv) \hspace{2cm} \bC \bar\psi(t,\xv)\bC^\dag = -i\psi^T(t,\xv)\aly
\eeq
For a meson state to be an eigenstate of charge conjugation,
\beq\label{A14}
\bC\ket{M,\Pv} = \int d\xv_1\,d\xv_2\,\bar\psi(\xv_1)e^{i\Pv\cdot(\xv_1+\xv_2)/2}\aly\big[\Phi^{(\Pv)}(\xv_2- \xv_1)\big]^T\aly \psi(\xv_2)\ket{0} = \eta_C \ket{M,\Pv}
\eeq
its wave function should satisfy
\beq\label{A15}
\aly\big[\Phi^{(\Pv)}(-\xv)\big]^T\aly = \eta_C \Phi^{(\Pv)}(\xv)  \hspace{2cm} (\eta_C = \pm 1)
\eeq

\section{Derivation of the alternative form \eq{eVI7} of the BSE \eq{eVI6} \label{B}}

We make use of commutator identities such as,
\beqa
\com{A}{BC}&=&\com{A}{B}C+B\com{A}{C} \label{eB1} \\[2mm]
\acom{A}{BC}&=&\com{A}{B}C+B\acom{A}{C} = \acom{A}{B}C-B\com{A}{C} \label{eB2} \\[2mm]
\acom{A}{\acom{B}{C}} &=& -\com{B}{\com{A}{C}} \hspace{1.15cm} {\rm when}\ \ \acomb{A}{B}=0 \label{eB3} \\[2mm]
\acom{A}{\com{B}{C}} &=& -\acom{B}{\com{A}{C}} \hspace{1cm} {\rm when}\ \ \acomb{A}{B}=0 \label{eB4} \\[2mm]
\com{A}{\acom{B}{C}} &=& -\com{B}{\acom{A}{C}} \hspace{1cm} {\rm when}\ \ \acomb{A}{B}=0 \label{eB5} \\[2mm]
\acom{A}{\com{A}{C}} &=& \com{A}{\acom{A}{C}}=\com{A^2}{C}  \label{eB6} \\[2mm]
\acom{A}{\acom{A}{C}} &=& 2A\acom{A}{C} \hspace{1.5cm} {\rm when}\ \ A^2=1 \\[2mm] \label{eB7}
\com{A}{\com{A}{C}} &=& 2A\com{A}{C} \hspace{1.65cm} {\rm when}\ \ A^2=1  \label{eB8}
\eeqa

Taking the commutator $\com{i\nv\cdot \alv}{\rm BSE}$ of the bound state equation \eq{eVI6} gives
\begin{align} \label{eB9}
\com{i\nv\cdot \alv}{(E-V)\wfp} &= \com{i\nv\cdot \alv}{\acomb{i\nv\cdot\alv}{\wfp}}
-\halft\com{i\nv\cdot \alv}{\comb{\Pv\cdot\alv}{\wfp}}+m\com{i\nv\cdot \alv}{\comb{\gz}{\wfp}}
\end{align}
The first term on the rhs. vanishes due to the commutator identity \eq{eB6}, when we recall that $\nv$ in the BSE always operates on $\wfp$.
The identity \eq{eB3} implies for the third term on the rhs. of \eq{eB9},
\begin{align} \label{eB10}
m\com{i\nv\cdot \alv}{\comb{\gz}{\wfp}} = -m\acom{\gz}{\acomb{i\nv\cdot \alv}{\wfp}}
\end{align}
Using the original BSE \eq{eVI6} on the rhs. we get
\begin{align} \label{eB11} 
m\com{i\nv\cdot \alv}{\comb{\gz}{\wfp}} &= -m\acom{\gz}{\halft \com{\Pv\cdot\alv}{\wfp}}+m^2\acom{\gz}{\com{\gz}{\wfp}}-m\acom{\gz}{(E-V)\wfp} \nn\crt
&\hspace{-1cm}= \halft m \acom{\Pv\cdot\alv}{\com{\gz}{\wfp}}-m(E-V)\acom{\gz}{\wfp} \nn\crt
&\hspace{-1cm}= \halft\acom{\Pv\cdot\alv}{-\acomb{i\nv\cdot\alv}{\wfp}+\halft\comb{\Pv\cdot\alv}{\wfp}+(E-V)\wfp}-m(E-V)\acom{\gz}{\wfp}
\end{align}
where we used \eq{eB4}, \eq{eB6} and in the last step expressed $m\com{\gz}{\Phi_P}$ using the BSE \eq{eVI6}. The second term on the rhs. of \eq{eB11} vanishes according to \eq{eB6}. Inserting this result in \eq{eB9} we have
\begin{align} \label{eB12} \hspace{-.4cm}
\com{i\nv\cdot \alv}{(E-V)\wfp} &= \nn\crt
& \hspace{-3.6cm}-\halft\com{i\nv\cdot \alv}{\com{\Pv\cdot\alv}{\wfp}}-\halft\acom{\Pv\cdot\alv}{\acom{i\nv\cdot \alv}{\wfp}} + \halft(E-V)\com{\Pv\cdot\alv}{\wfp}-m(E-V)\acom{\gz}{\wfp}
\end{align}
The sum of the first two terms on the rhs. simplifies. With $\nv\cdot \alv=\alpha^i\partial_i$ and $\Pv\cdot \alv=P^j\alpha^j$,
\begin{align} \label{eB13}
\com{\alpha^i}{\com{\alpha^j}{\partial_i\wfp}} &= \alpha^i(\alpha^j\partial_i\wfp- \partial_i\wfp\alpha^j)-(\alpha^j\partial_i\wfp- \partial_i\wfp\alpha^j)\alpha^i
\nn\crt
\acom{\alpha^j}{\acom{\alpha^i}{\partial_i\wfp}} &= \alpha^j(\alpha^i\partial_i\wfp+ \partial_i\wfp\alpha^i)+(\alpha^i\partial_i\wfp+ \partial_i\wfp\alpha^i)\alpha^j
\end{align}
so that
\begin{align} \label{eB14}
\com{\alpha^i}{\com{\alpha^j}{\partial_i\wfp}}+\acom{\alpha^j}{\acom{\alpha^i}{\partial_i\wfp}} &= (\alpha^i\alpha^j+\alpha^j\alpha^i)\partial_i\wfp+ \partial_i\wfp(\alpha^j\alpha^i+\alpha^i\alpha^j) = 4\partial_j\wfp
\end{align}
Using this in \eq{eB12} and dividing by $E-V$ gives
\begin{align} \label{eB15}
\inv{E-V}\com{i\nv\cdot \alv}{(E-V)\wfp}-\halft\acom{\Pv\cdot\alv}{\wfp}+m\acom{\gz}{\wfp} = -\frac{2i}{E-V}\Pv\cdot\nv\wfp
\end{align}
For a linear potential $i\nv\cdot\alv\, V'|\xv|= iV'\alv\cdot\xv/r$, where $r=|\xv|$. Bringing this derivative to the rhs. in \eq{eB15},
\begin{align} \label{eB16}
\com{i\nv\cdot \alv}{\wfp}-\halft\acom{\Pv\cdot\alv}{\wfp}+m\acom{\gz}{\wfp} = \inv{E-V}\Big(-2i\Pv\cdot\nv\wfp+\frac{V'}{r}\com{i\alv\cdot\xv}{\wfp}\Big)
\end{align}
The lhs. is now the same as in the original BSE \eq{eVI6}, with commutators and anticommutators interchanged. Adding and subtracting the two equations and dividing by $E-V$ we get equations \eq{eVI7}.

\bibliography{181221_refs.bib}

\end{document}